\documentclass[lettersize,journal]{IEEEtran}
\usepackage{aliases}
\usepackage{amsmath}
\usepackage{tikz}
\usepackage{xspace}
\usepackage{amsthm}
\usepackage{booktabs} 
\usepackage{algorithm}
\usepackage[noend]{algpseudocode}
\usepackage{amssymb}
\usepackage{amsthm}
\usepackage{subcaption}
\usepackage{tabularx}
\usepackage{pifont}
\usepackage[group-separator={,},per-mode=symbol,binary-units=true]{siunitx}
\usepackage[normalem]{ulem}
\useunder{\uline}{\ul}{}
\usepackage{booktabs}
\usepackage{makecell}

\usepackage[T1]{fontenc}        %
\usepackage[utf8x]{inputenc}

\usepackage{amsmath,amssymb,amsthm,array,xcolor,colortbl,graphicx,multirow}
\usepackage{microtype}
\usepackage{comment}
\usepackage{balance}
\usepackage{stmaryrd}
\usepackage{import}
\usepackage{relsize}
\usepackage{scalefnt}
\usepackage{xspace}
\usepackage{csquotes}
\usepackage{epsfig,graphicx}  %
\usepackage{graphicx, amssymb}
\usepackage{footnote}
\usepackage{tikz}
\usetikzlibrary{shapes}
\usepackage{array,xcolor,colortbl,graphicx,multirow}
\usepackage[multiple]{footmisc}
\allowdisplaybreaks[3]

\usepackage{cite}

\newcommand{\figcap}[1]{\caption{#1}}
\newcommand{\sfigcap}[1]{\subcaption{#1}}
\newcommand{\tabcap}[1]{\vspace*{5pt}\caption{\textit{#1}}}

\algnewcommand{\LeftComment}[1]{\State  \(\triangleright\) #1 \hfill~}

\usepackage{etoolbox}
\usepackage[pdfpagelabels=true,hidelinks]{hyperref}

\usepackage[alignedforall]{lpform}
\usepackage{epigraph}

\newcommand{\uns}[1]{\SI{#1}{\nano\second}}
\newcommand{\umus}[1]{\SI{#1}{\micro\second}}
\newcommand{\us}[1]{\SI{#1}{\second}}
\newcommand{\uGB}[1]{\SI{#1}{\giga\byte}}

\newcommand{\uGbps}[1]{\SI{#1}{Gbps}}
\newcommand{\uMB}[1]{\SI{#1}{\mega\byte}}
\newcommand{\uKB}[1]{\SI{#1}{KB}}

\begin{document}

\title{\LARGE \bf Unlocking Diversity of Fast-Switched\\Optical Data Center Networks with Unified Routing}

\author{Jialong Li, Federico De Marchi, Yiming Lei, Raj Joshi, Balakrishnan Chandrasekaran, Yiting Xia
\thanks{Manuscript received XXX XX, XXX; revised XXX XX, XXXX.}}

\markboth{Transactions on Networking,~Vol.~XX, No.~XX, October~2024}%
{Shell \MakeLowercase{\textit{et al.}}: A Sample Article Using IEEEtran.cls for IEEE Journals}

\maketitle

\begin{abstract}

Optical data center networks (DCNs) are emerging as a promising solution for cloud infrastructure in the post-Moore's Law era, particularly with the advent of ``fast-switched'' optical architectures capable of circuit reconfiguration at microsecond or even nanosecond scales. However, frequent reconfiguration of optical circuits introduces a unique challenge: in-flight packets risk loss during these transitions, hindering the deployment of many mature optical hardware designs due to the lack of suitable routing solutions.

In this paper, we present \textbf{U}nified \textbf{R}outing for \textbf{O}ptical networks (\name), a general routing framework designed to support fast-switched optical DCNs across various hardware architectures. \name combines theoretical modeling of this novel routing problem with practical implementation on programmable switches, enabling precise, time-based packet transmission. Our prototype on Intel Tofino2 switches achieves a minimum circuit duration of $\umus{2}$, ensuring end-to-end, loss-free application performance. Large-scale simulations using production DCN traffic validate \name's generality across different hardware configurations, demonstrating its effectiveness and efficient system resource utilization.

\end{abstract}

\begin{IEEEkeywords}
Data center networks, routing, optical data center networks, programmable switches.
\end{IEEEkeywords}

\section{Introduction}\label{sec:intro}%

\begin{figure}[t]
    \centering
    \includegraphics[width=0.99\columnwidth]{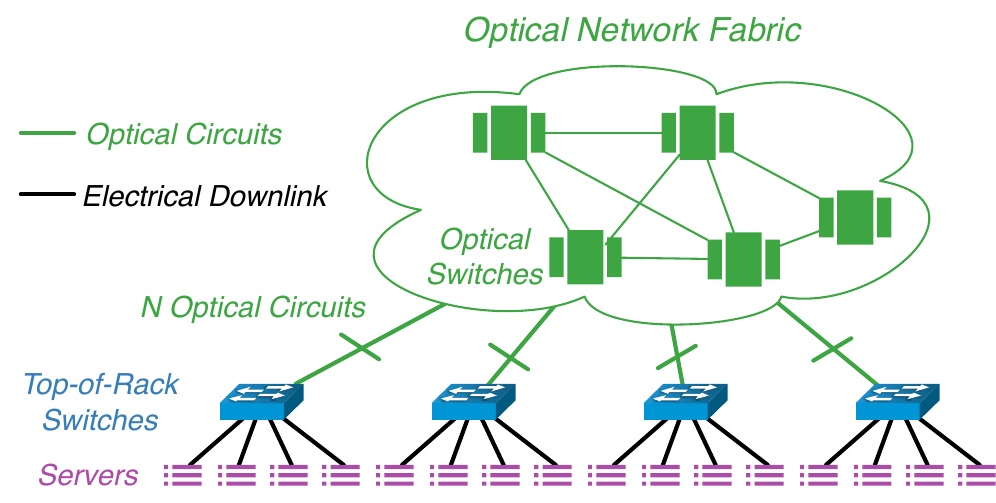}
    \figcap{Illustration of an optical \dcn.
    } \vspace{-0.1in}
    \label{fig:arch}
\end{figure}

\IEEEPARstart{I}{n} the post-Moore’s law era for merchant silicon, optical data center networks (\dcns) are emerging as the future cloud network infrastructure for their power, cost, and bandwidth advantages. They function in a fundamentally different way compared to traditional \dcns. As illustrated in Fig.~\ref{fig:arch}, an optical \dcn creates reconfigurable optical circuits between top-of-rack switch (ToR) pairs through a set of optical circuit switches (\ocses). Each circuit lasts for a fixed interval of time, called a ``time slice,'' and the network topology changes over time as the \ocses reconfigure the circuits.

Optical \dcns were initially designed to be \textit{slow-switched}, using coarsely reconfigured \ocses with millisecond-scale reconfiguration delays to offer time slices lasting seconds or longer. This design aims to achieve high throughput for bulk data transfers, commonly referred to as ``elephant flows''~\cite{Helios, cThrough, OSA, JupiterEvolving, Flat-tree, WaveCube, MegaSwitch, Firefly, OmniSwitch}. With the advancement of microsecond- and nanosecond-scale optical switching technologies, however, the focus has increasingly shifted to \textit{fast-switched} designs. They leverage microsecond- or even sub-microsecond-scale time slices, enabled by fine-grained \ocs reconfigurations, to also serve latency-sensitive traffic, known as ``mice flows''~\cite{Mordia, RotorNet, Opera, Sirius, PULSE, Modoru, RAMP, HWS-TDMA-1, POTORI, Flex-LIONS, ProjecToR}.

The ultimate goal of fast-switched optical \dcns is to achieve packet-granularity reconfiguration like electrical packet switches~\cite{Sirius, PULSE, ORN, ORN_2}, and the pioneering work of Sirius has demonstrated feasibility of the hardware architecture~\cite{Sirius}.
Nevertheless, circuit reconfigurations at per-packet frequency,
i.e., with nanosecond-scale time slices,
is extremely disruptive to the existing network stack.
Consequently, this line of work requires a complete redesign of the network from scratch, making end-to-end deployment a long-term endeavor.

Decades of optics research has made available a wide variety of optical technologies spanning a full spectrum of reconfiguration delays. Numerous network architectures have been proposed to explore different time slice durations as transitional solutions to realizing the above vision (see~Table~\ref{tab:architecutures}).
Unfortunately, networked systems for these hardware architectures seriously lag behind, with a few customized software systems tailored to specific design points of optical hardware. This co-design strategy impedes the development cycle and limits diversity of fast-switched optical \dcns.

In this paper, we propose \textit{a unified routing solution} --- \textbf{U}nified \textbf{R}outing for \textbf{O}ptical networks (\name\footnote{Pronounced as ``Euro'' and inspired by the fact that Euro unified currencies in the European Union.}) --- to change this \textit{status quo}, because routing is the most critical network function directly interfacing with the optical hardware. We borrow the philosophy of IP as the narrow waist in the traditional network stack, using a simple unified routing design to unlock diversity of optical architectures as well as potential upper-layer protocols. We believe this strategy
is a promising near-term solution to interoperability and independent evolution of optical and networking technologies.

The challenge, though, lies in the short time slices.
In fast-switched optical \dcns, a time slice
can be shorter than a packet's one-way delay (\owd) over the routing path, from the source ToR to the destination ToR, possibly traversing intermediate ToRs. In-flight packets may encounter circuit reconfiguration and risk being lost. To prevent losses, paths must be meticulously planned ahead of time; and packets must act precisely according to the plan, stopping at the affected intermediate ToR before reconfiguration occurs and waiting until a new circuit is established to continue routing.

\begin{table}[tbp]
\footnotesize
\centering
\tabcap{Architectures utilizing different optical hardware, those originally incorporate a networked system are \underline{underlined}, those supported by \name are highlighted in \textcolor{blue}{blue}.}
\resizebox{\columnwidth}{!}{%
\begin{tabular}{ccc}
\toprule
Architecture & Reconfiguration delay & Time slice duration  \\
\midrule
PULSE~\cite{PULSE}        & 500 $ps$              & 20 $ns$              \\
\underline{Sirius~\cite{Sirius}}       & 1 $ns$                & 40 $ns$              \\
\textcolor{blue}{Modoru}~\cite{Modoru}       & 10 $ns$               & $\mu s$-level \\
\textcolor{blue}{RAMP}~\cite{RAMP}         & 10 $ns$               & $\mu s$-level        \\
\textcolor{blue}{HWS-TDMA}~\cite{HWS-TDMA-1}     & 200 $ns$              & 2 $\mu s$            \\
\textcolor{blue}{POTORI}~\cite{POTORI}       & 2 $\mu s$             & 10 $\mu s$           \\
\textcolor{blue}{Flex-LIONS}~\cite{Flex-LIONS}   & 10 $\mu s$            & 100 $\mu s$          \\
\underline{\textcolor{blue}{ProjecToR}~\cite{ProjecToR}}    & 12 $\mu s$            & 120 $\mu s$          \\
\underline{\textcolor{blue}{RotorNet}~\cite{RotorNet} / \textcolor{blue}{Opera}~\cite{Opera}}     & 20 $\mu s$            & 200 $\mu s$          \\
\bottomrule
\end{tabular}%
}
\label{tab:architecutures}
\end{table}

Sub-\owd time slices break the basic assumption in static networks that packets follow \textit{continuous paths} without interruptions. As shown in Table~\ref{tab:architecutures}, this challenge motivated most software systems to be designed for super-\owd time slices.
Particularly, as an example of topology-routing co-design, Opera~\cite{Opera} constrains the time slice duration to be much larger than \owd by partially reconfiguring the topology in each time slice and keeping most circuits stable. This approach effectively ensures all packets to go through continuous paths, but at the cost of dilated paths, causing high latency for mice flows.

Since \name is a unified solution general to different slice durations, we combine theory and practice to tackle the challenge. From a theoretical perspective, we formulate this new routing problem involving ``waiting'' at intermediate ToRs,
and redefine the routing latency to incorporate circuit-waiting delays at ToRs. Fast-switched optical \dcns are equipped with a cyclic schedule of circuit connections known \textit{a priori}, due to the high time precision of operation required. Leveraging this fact, we design an \textit{offline} routing algorithm to compute paths with the objective of minimizing (the redefined) latency.

On the practical side, ``waiting'' implies buffering packets on ToRs and controlling their behaviors on a fine time basis, which is often deemed impossible. We, however, leverage innovations in programmable switches to realize these functionalities. Programmable switches have demonstrated support for temporal buffering for a small number of packets~\cite{SQR,goswami2020parking}, which is sufficient for sub-\owd time slices. Time-based packet control can be achieved with the on-chip packet generator and queue pausing/resuming feature~\cite{lee2020tofino2}, where we create control packets by the packet generator at nanosecond precision to enable/disable packet transmission.

We implemented \name on Intel Tofino2 switches, which supports a minimum time slice of $\umus{2}$.
As shown in Table~\ref{tab:architecutures}, \name uniquely supports sub-\owd time slices and extends to cover super-\owd ones, as our routing algorithm reduces to $k$-shortest path routing in that range. We also showed on the testbed that with varying time slice durations from $\umus{2}$ to $\umus{50}$, applications run end-to-end without packet losses, with negligible differences in flow completion times (\fcts).

In our large-scale simulations with production \dcn traffic, \name exhibits 10.0\%-14.8\% reduction of path length and 1.4$\times$-12.8$\times$ lower \fcts than Opera. The \name algorithm generalizes to arbitrary time slices. Under the hypothetical case of packet-granularity time slices, imagining future support from the ToRs, \name achieves comparable lower-bound \fcts with Sirius~\cite{Sirius} and its variants~\cite{ORN, ORN_2}. In all our experiments from $\umus{2}$ to $\umus{300}$ time slices, \name consumes at most $\uKB{410}$ of packet buffer and 23 queues per egress port,
significantly below the capacity limit of commodity switch ASICs~\cite{BFC,SQR}.

\textit{[This work does not raise any ethical issues.]}

\section{Background} \label{sec:background}%

\subsection{Fast-Switched Optical \dcns}\label{sec:rdcn}

An optical \dcn fabric (Fig.~\ref{fig:arch}) uses \ocses to construct reconfigurable optical circuits between different ToR pairs. Optical \dcns can be classified into two categories:
traffic-aware and traffic-oblivious.
Traffic-aware optical \dcns estimate traffic demands to configure their circuits on demand~\cite{MegaSwitch, Mordia, OSA, ProjecToR, Flat-tree, Firefly, Gemini}.
For fast-switched optical \dcns studied in this paper,
estimating traffic demands in a timely and precise manner poses a considerable challenge.
Therefore, they adopt the traffic-oblivious design, which
constructs the optical circuits in a predefined way, regardless of traffic patterns~\cite{Opera, RotorNet, Sirius, Shoal, ORN}. The \ocses continuously change the circuits, and each circuit lasts for a fixed interval of time, called a \textit{time slice}. The switch repeats the schedule continuously and guarantees that each ToR pair is assigned at least one circuit per repetition (or \textit{cycle}). A sequence of time slices within \textit{one cycle}, each connecting some subset of ToR pairs, constitutes an \textit{optical schedule}. Typically, a \textit{cycle} consists of several tens of \textit{time slices}, and a \textit{time slice} spans from sub-microseconds to milliseconds. In each time slice, the optical \dcn functions as a static graph, essentially forming a sequence of time-varying graphs that cyclically repeat.

\subsection{Routing in Fast-Switched Optical \dcn}\label{sec:prior_routing}

The most intuitive way of routing in an optical \dcn is through direct circuits, but the down side is long latency waiting for the circuit to appear. As a result, multi-hop routing schemes via intermediate ToRs has become prevalent to leverage more readily available circuits. Because most routing solutions are coupled with the underlying architecture ($\S$\ref{sec:intro}), we use the architecture name to refer to the adopted routing approach.

\para{Sirius.}
Sirius~\cite{Sirius} adopts cutting-edge customized hardware that allows for \textit{packet-granularity} time slice duration, meaning that \textit{only one packet} is sent out in each time slice.
Sirius uses Valiant Load Balancing (\vlb) routing to support arbitrary traffic patterns.
\vlb is a two-phase routing scheme which is roughly equivalent to packet spraying in optical \dcns.
In \textit{phase~1}, a source \tor randomly sprays packets in their first hop to directly connected intermediate \tors.
Then in \textit{phase~2}, packets wait at intermediate \tors to be forwarded to their destination.
While generally \vlb waiting in \textit{phase~2} would produce high tail latency for mice flows, Sirius can adopt it with little impact thanks to its packet-granularity optical schedule.
While Sirius has been demonstrated as a small FPGA prototype, the feasibility of packet-granularity time slices is still unclear for actual, large-scale deployments.

\para{\orn.}
Vandermonde Bases Scheme (\orn)~\cite{ORN, ORN_2} is a recent theoretical contribution that builds upon the same architectural assumptions as Sirius (i.e., packet-granularity schedule plus \vlb routing), and that provides a generalization of the single-dimensional round-robin schedule used in Sirius.
The \orn generalization allows for a multi-dimensional round-robin schedule design, which is tunable by setting a parameter $h$.
$h$ sets the number of round-robins a packet goes through to reach its destination, where higher $h$ corresponds to lower latency but higher bandwidth expense.
To clarify, the single-dimensional schedule can be seen as a single-dimensional hypercube, while the multi-dimensional schedule as an $h$-dimensional hypercube. It is safe to assume \orn would need no less engineering effort than Sirius.

\para{Opera.}
Opera~\cite{Opera} builds upon super-\owd time slices.
As exclusively relying on \vlb would result in unfeasible latency for mice flows at this time slice granularity, Opera judiciously designs its optical schedule so that at any point in time \tors offer always-available, continuous paths over a time-varying expander graph.
In order to reliably support these paths, Opera has to make compromises in its architecture.
First of all, as a way to keep the expander always connected, only a restricted number of circuits is allowed to reconfigure at once, and paths have to be more redundant than in an optimal expander.
Secondly, a time slice in Opera must be held for at least a worst-case OWD in order to guarantee packets in transit do not cross a reconfiguring circuit.
This OWD is kept at a reasonable value by bounding the maximum buffer sizes to a strict amount, i.e., the congestion threshold in the coupled NDP~\cite{NDP} protocol.
In Opera,
elephant flows are routed by high-throughput but high-latency \vlb routing, and mice flows are sent through continuous shortest-path routing over the always-available paths.
The cutoff between the two flow classes is 15\,MB.

\section{\name Algorithm}\label{sec:design}%

In this section, we define the routing problem for \hoho ($\S$\ref{sec:formulation}) and describe the \hoho algorithm design ($\S$\ref{sec:alg_design}). We prove three critical properties of the algorithm ($\S$\ref{sec:properties}), which serve as the theoretical foundation for
the ToR system ($\S$\ref{sec:implementation}).

\subsection{Problem Formulation}\label{sec:formulation}

We model an optical \dcn as a time-varying graph $G=(V, E, T)$, where the vertices ($V$) denote the ToRs and the edges ($E$) represent the {\em time-dependent} optical circuits connecting those ToRs.
$T$ is the optical schedule; for each edge $e \in E$, $T_e = \{t_i, t_j, ..., t_k\}$ represents the time slices (of fixed durations) during which $e$ exists.
We specify a routing path from a source ($src$) to a destination ($dst$) node in this graph by $p(src, dst, t_{start})$, where $t_{start}$ is the time slice when that path is available.
We use this path for transmitting packets that arrive at $src$ in this time slice and destined for $dst$.
If the path is composed of more than one segment, i.e., constituting one or more intermediate hops (i.e., ToRs), we may buffer the packets at each hop depending on when the subsequent segment of the path becomes available.
Packets may, hence, reach $dst$ at a time ($t_{end}$) later than $t_{start}$.
Our objective, naturally, is to minimize the latency of the path traversed by the packets.

\begin{equation}\label{eqn:latency}
   lat(p)= (t_{end} - t_{start} + 1) \times u
\end{equation}

\para{Path latency.}
We define latency of a path in an optical \dcn as above, by the number of elapsed time slices during routing multiplying the time slice duration ($u$) (Eqn.~\ref{eqn:latency}).
With this simple definition, we can use an offline algorithm for computing the low-latency paths between every node pair based on the \dcn's pre-determined optical schedule.
Eqn.~\ref{eqn:latency} captures, nevertheless, the effect of circuit reconfigurations on path latencies, especially for sub-\owd time slices.
Traditional routing schemes for optical \dcns ($\S$\ref{sec:prior_routing}), in contrast, do not account for the implications of reconfigurations:
They assume either relatively long time slices, which enable packets to be routed strictly within one time slice, or ultra-short packet-granularity time slices, where the routing latency across multiple time slices is negligible.
Our latency definition also takes the transmission and propagation delays---which are at most on the order of hundreds of nanoseconds\footnote{The transmission delay for a 1500\,B packet under $\uGbps{100}$ is $\uns{120}$, and the propagation delay is $\uns{5}$ per meter of cable; $lat(p)$ is strictly larger than   \owd of packet delivery, which is at least a few microseconds in \dcns.} and, hence, substantially smaller than $lat(p)$---into account.

\begin{figure}[tb]
    \centering%
    \includegraphics[width=0.60\linewidth]
    {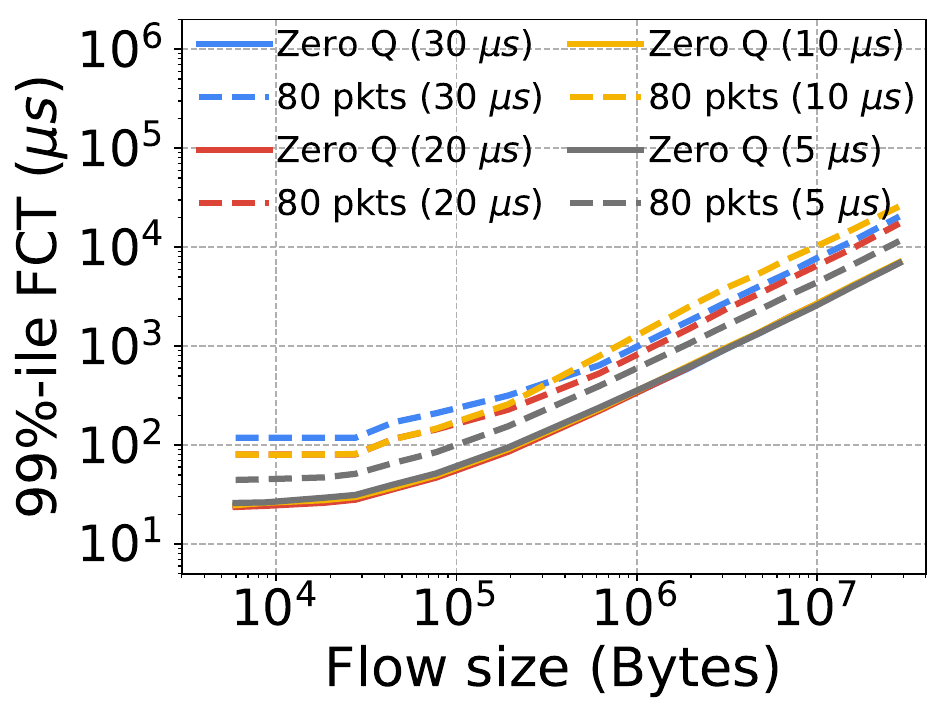}
    \caption{\hoho tail \fcts assuming empty queues vs. 80 packets per queue with 30\% ToR-to-ToR link utilization. 
    }
    \label{fig:fct_fix_q}
    \vspace{-0.00in}
\end{figure}

\para{What of queuing delays?} We do not explicitly model queuing delays, following the convention of classic routing algorithms.
Our decision was also influenced by the technical challenges in measuring queuing delays at microsecond granularity in real time.
Most importantly, we observe that under production \dcn traffic, considering queuing delays, even with shallow queues, result in poor \fcts.
Prior work have shown, for instance, that the median link utilization in production \dcns is 10\%-20\%~\cite{FB_traffic}, with 80\% of the time below 10\%~\cite{zhang2017high}.
Even with traffic loads higher than production \dcns{}, the current best practice of factoring in queuing delays---estimated from worst-case queue occupancies~\cite{Opera}---precipitates in overestimating delays; it offers, hence, significantly larger FCTs than those obtained by assuming zero queuing delays.

We simulate this scenario in Fig.~\ref{fig:fct_fix_q} by generating traffic on a $\uGbps{100}$ 108-ToR optical \dcn using the web search trace~\cite{VL2}. The ToR-to-ToR link utilization is set to a high load of 30\%. Further simulation details are available in $\S$\ref{sec:setup}.
The best practice for incorporating queuing delays into routing of optical \dcns is Opera~\cite{Opera}. It adopts NDP as the transport protocol and takes the NDP congestion threshold as the worst-case queue occupancy. This threshold, e.g., 80 packets per queue at $\uGbps{100}$, is carefully chosen for a low-latency transport protocol to balance shallow queues and high throughput. However, we argue that it is an overestimation for routing designs in optical \dcns, given the typically low \dcn traffic loads.

As illustrated in Fig.~\ref{fig:fct_fix_q}, even under higher traffic loads than typical production \dcns, assuming empty queues achieves 1.7$\times$ to 4.9$\times$ lower 99\textsuperscript{th} percentile \fcts than factoring in fixed queuing delays as done in Opera, 
across a wide range of time slice durations.
Overestimating queuing delays cause packets to act conservatively. They may mistakenly bypass feasible circuits believing they cannot complete transmission and instead choose later circuits as safer options.

\para{Takeaways.}
Our objective in \hoho is to minimize the latency in Eqn.~\ref{eqn:latency} under two assumptions:
(a) packets always arrive at the beginning of a time slice and (b) there is no queuing delay at the ToRs.
These assumptions enable us to decouple the routing design into a \textit{static} offline routing algorithm and a \textit{run-time} on-switch system.
The offline algorithm computes the paths, comprising zero or more intermediate hops (i.e., ToRs), based on the network's pre-determined optical schedule ($\S$\ref{sec:rdcn}); it captures the implications of reconfigurations for path latencies, since reconfigurations may cause packets to wait at an intermediate hop until the next segment of the path becomes available.
The run-time on-switch system, in contrast, copes with deviations in a packet's actual arrival time (due to, for instance, queuing delays) and re-routes the packet if it misses its scheduled time slice.

\subsection{Algorithm Design}\label{sec:alg_design}

\begin{figure}[t]
    \centering
    \includegraphics[width=1.0\columnwidth]{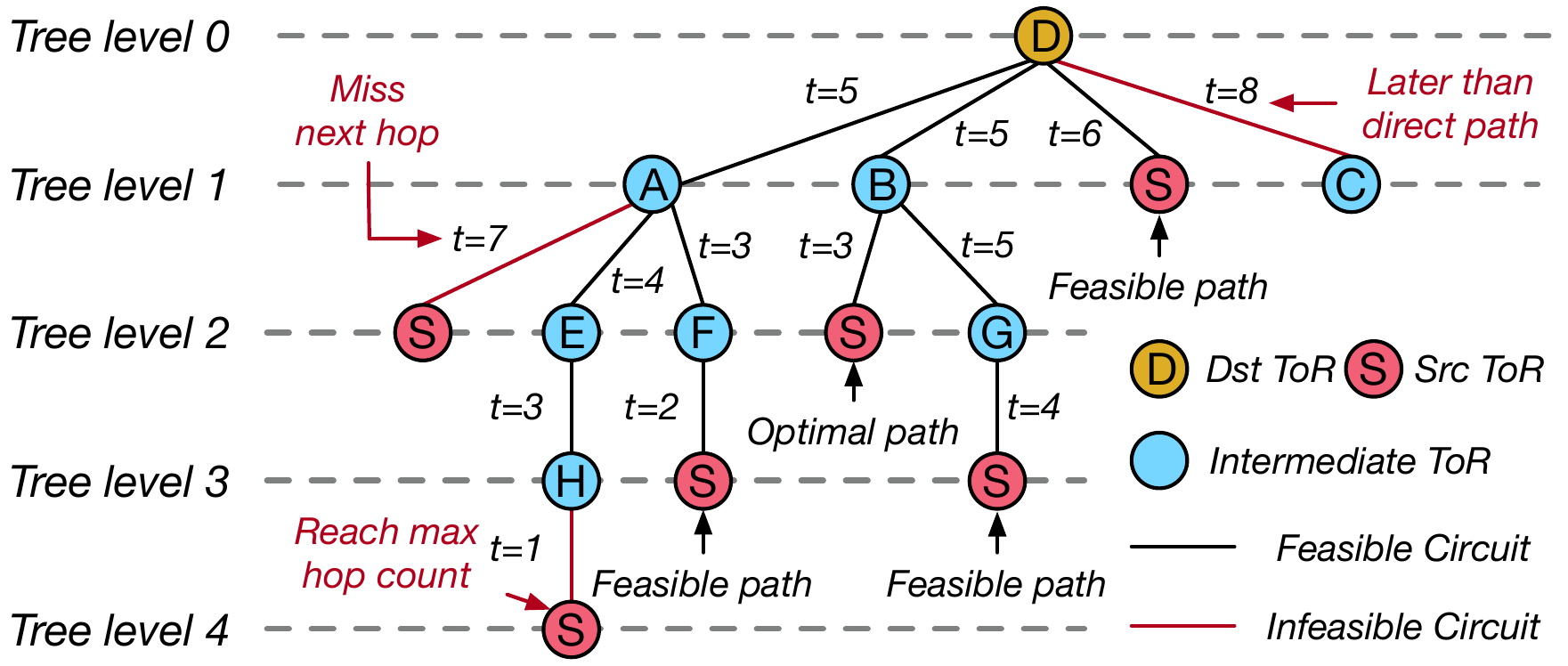}
    \figcap{Illustration of the \hoho backtracking algorithm. The packet arrival time slice is $t=0$. Optical circuits are denoted as edges with their available time slices annotated on top.
    The destination calls \textsc{Routing} to find the earliest last hops ($A$ and $B$ with $t=5$), which then call \textsc{Subpath} to find the shortest feasible path through them from the source ($S\rightarrow B\rightarrow D$). Paths violating various constraints (see explanations in red) are pruned from the backtracking search.
    }
    \label{fig:hoho_example}
\end{figure}

We design the \name algorithm to solve the routing problem for optical \dcns defined in $\S$\ref{sec:formulation}. We explain the algorithm with the example in Fig.~\ref{fig:hoho_example}.

For a packet that arrives at the source ToR $src$ in time slice $t$, the time to reach the destination ToR $dst$ is solely determined by the time slice of the circuit connecting the last-hop ToR $r$ to the destination $dst$. This is because, if we view optical circuits in an optical \dcn as ``buses'' and packets as ``passengers'' to be transported over these circuits, the arrival time at the destination depends on when the ``bus'' from the last ``stop'' departs for the destination. This is true irrespective of the number of ``transitions'' the ``passenger'' undergoes, as long as they catch the last ``bus''.

Following this intuition, we design a \textit{backtracking} algorithm for \name that comprises two procedures: \textsc{Routing} and \textsc{Subpath}, which correspond to the two steps, respectively. In Fig.~\ref{fig:hoho_example}, we illustrate the backtracking search tree originating from the destination ToR. The edges in the tree represent the circuits between the ToRs. The time slice when each circuit is available is indicated above the edge.

Therefore, in \name, finding the \textit{fastest} path to deliver the packet from $src$ in time slice $t$ to $dst$ with the minimum delay involves two steps. \textit{Step 1} is to identify the earliest ``bus'' at the last hop $r$ to the destination, given the ``bus'' (circuit) schedule. \textit{Step 2} is to plan an ``itinerary'' from $src$ to $r$ that meets the ``deadline'' of catching the next ``bus'' at each ``transition''. This means arriving either before or exactly at the time when the next ``bus'' departs, i.e., earlier than or precisely within the time slice scheduled for the onward hop.

\begin{algorithm}[t]
  \small
  \caption{Unified Routing Algorithm}
  \label{alg:hoho}
  \begin{algorithmic}[1]
  \Require
	\Statex $M$ $ \leftarrow $ max hop count
	\Statex $src$, $dst $ $ \leftarrow $ source ToR, destination ToR
	\Statex $ t_{0} \leftarrow $ the packet arrival time slice at $src$
	\Statex $ t_{(s, d)} \leftarrow $ the earliest time slice when ToR $s$ and ToR $d$ are connected, where $t_{(s, d)} \geq t_0$ must hold. $ t_{(s, d)} $ is derived by the optical schedule $L$

   \LeftComment{Find the fastest path per ToR pair per time slice}
    \Procedure{Routing($src$, $dst$, $t_0$)}{}
        \State{Sort all ToRs by $t_{(r, dst)}$ in ascending order}
        \State{$path = \emptyset$, $min\_time = t_{(r_{[0]}, dst)}$, $min\_hop = \infty$}
        \For{each $r$ in ToRs}
            \If{$t_{(r, dst)} > min\_time$ \textbf{and} $path \neq \emptyset$}
                \State{\Return{$path$}}
            \EndIf
            \State{$min\_time = t_{(r, dst)}$}
            \State{$path' = $ \Call{Subpath}{$src$, $r$, $t_{(r, dst)}$, 1, $\{dst\}$}}
            \If{$path' \neq \emptyset$ \textbf{and} $|path'| < min\_hop$}
                \State{$path = path'$, $min\_hop = |path'|$}
            \EndIf
        \EndFor
        \State{\Return{$path$}}
    \EndProcedure
    
    \LeftComment{Find the shortest feasible subpath through an intermediate ToR}
    \Procedure{Subpath($src$, $r$, $t$, $level$, $subpath$)}{}
        \If{$level > M$}
            \State{\Return{$\emptyset$}}
        \EndIf
        \If{$t_{(src, r)} \leq t$}
            \State{\Return{$src + subpath$}}
        \EndIf
        \State{$feasible = \{ \}$}
        \For{each $r'$ in ToRs \textbf{and} $r'$ not in $subpath$}
            \If{$t_{(r', r)} \leq t$}
                \State{$p$ = \Call{Subpath}{$src$, $r'$, $t_{(r', r)}$, $level+1$, $r + subpath$}}
                \If{$p \neq \emptyset$}
                    \State{$p\rightarrow feasible$}
                \EndIf
            \EndIf
        \EndFor
        \State{\Return{$shortest(feasible)$ \textbf{or} $\emptyset$}}
    \EndProcedure
  \end{algorithmic}
\end{algorithm}

For a packet that arrives at the source ToR in a particular time slice, the \textsc{Routing} procedure finds the fastest optical path to the destination ToR.
It finds the last-hop ToR that provides the earliest arrival at the destination ToR, by sorting the time slices of all candidate ToRs connecting to this destination (\textit{line 2}).
For each candidate ToR, it calls the \textsc{Subpath} procedure to find a feasible sub-path from the source ToR (\textit{line 20}).
The procedure exits on finding the first valid path (\textit{line 6}) or when the search ends (\textit{line 11}).
Since each ToR pair is guaranteed a circuit in the optical schedule (refer~\S\ref{sec:rdcn}), \textsc{Routing} will always find a path---the direct path (S$\rightarrow$D in Fig.~\ref{fig:hoho_example}) in the worst case, if no faster path exists.
When multiple fastest paths (via A and B in Fig.~\ref{fig:hoho_example}) exist, the shortest path is chosen (\textit{lines 9-10}).

The \textsc{Subpath} procedure finds a feasible sub-path from the source ToR to an intermediate ToR recursively.
The procedure can terminate in two ways:
(i) when it fails (\textit{lines 13-14}) to find a path of length at most the maximum hop count, e.g., S$\rightarrow$C$\rightarrow$E$\rightarrow$A$\rightarrow$D in Fig.~\ref{fig:hoho_example},
or
(ii) when it finds a connection from the source ToR and its time slice can make the ``deadline'' for the next-hop transmission (\textit{lines 15-16}).
In Fig.~\ref{fig:hoho_example}, for instance, S$\rightarrow$B$\rightarrow$D meets this condition, as the time slice \texttt{t=3} for S$\rightarrow$B is earlier than the time slice \texttt{t=5} for B$\rightarrow$D, while S$\rightarrow$A$\rightarrow$D violates this condition.
No matter how many hops are traversed, the path must start from the source ToR.
So, a path is found if and only if the source ToR is directly connected to the current intermediate ToR.
Otherwise, \textsc{Subpath} calls itself to search onward to other intermediate ToRs not already in the sub-path and finds feasible sub-paths that constantly meet ``deadlines'' (\textit{lines 18-22}).
If \textsc{Subpath} find multiple feasible sub-paths, we select the shortest one (\textit{line 23}).
In Fig.~\ref{fig:hoho_example},
S$\rightarrow$B$\rightarrow$D is chosen ultimately because it is shorter than the other feasible path S$\rightarrow$G$\rightarrow$B$\rightarrow$D.

The time complexity of \hoho algorithm depends on the number of ToRs, $N$, and the maximum hop count, $M$. Naively, the time complexity is $O(N^{M})$ since each node at the current tree level needs to check $N$ nodes at the next level. In practice, however, half of the nodes are filtered out at each level, reducing the number of nodes at each subsequent level by half, i.e., $N/2$, $N/4$, etc. This reduction continues, resulting in a time complexity of $1\times N/2 \times N/4 \times ... \times N/2^{M}$. Consequently, the overall time complexity is effectively reduced to $O(N^{M}/2^{M^2+M})$. This polynomial time algorithm completes computation in $\us{55}$ for our simulated 108-ToR optical \dcn ($\S$\ref{sec:evaluation}). \hoho applies to packet-granularity time slices. When the time slice duration exceeds the \owd, the latency definition in Eqn.~\ref{eqn:latency} simplifies to the constant value of the time slice duration.
In this case, \hoho effectively reduces to shortest-path routing, as all paths have equivalent latencies.

\subsection{Algorithm Properties}\label{sec:properties}

Below, we present and prove \hoho's three properties that are essential for implementing it on programmable switches.

\vspace{0.5em}\noindent\textbf{\textit{Property 1: The \name algorithm is optimal: the chosen path is the shortest that leads to the minimal latency.}}

\begin{proof}
Let $p$ be the selected path whose last-hop ToR to $dst$ is $r$ and path length is $l$. The time slice of the optical connection between $r$ and $dst$ is $t$. If there exists a better path $\hat{p}$ from $src$ to $dst$ with the last-hop ToR $\hat{r}$ at slice $\hat{t}$ and the path length is $\hat{l}$, then either $t$ > $\hat{t}$, or $t$ = $\hat{t}$ and $l > \hat{l}$. We prove by contradiction:

Case I: $t$ > $\hat{t}$. In \textsc{Routing}, last-hop ToRs are traversed by their time slices to $dst$ ascendingly. So, $\hat{p}$ must be found earlier than $p$, which is a contradiction.

Case II: $t$ = $\hat{t}$ and $l > \hat{l}$. When \textsc{Routing} breaks the tie on the same time slice at the last hop, $\hat{p}$ would overwrite $p$ and be chosen,
which is a contradiction.

\par \vspace{-\baselineskip}
\qedhere
\end{proof}

\name produces full paths, including every hop along the way, but the routing lookup on each intermediate ToR is based only on the immediate next hop.
Now, we prove that this implementation preserves the optimal paths.

\vspace{0.5em}\noindent\textbf{\textit{Property 2: Per-hop lookups yield the optimal path.}}

\begin{proof}

Let $p$ be the selected path whose first-hop ToR from $src$ is $r$, last-hop ToR to $dst$ is $r'$, the optical connection between $src$ and $r$ is at time slice $t$, the connection between $r'$ and $dst$ is at slice $t'$, and the path length is $l$. The residual path from $r$ to $dst$ is $p' = p - src$, the arrival time at $r$ is $t$, and the path length is $l' = l - 1$. We prove $p'$ is an optimal path for \textsc{Routing}($r$, $dst$, $t$). 

If there exists a better path $\hat{p'}$ than $p'$ from $r$ to $dst$ at slice $t$, whose last-hop ToR to $dst$ is $\hat{r'}$, the optical connection between $\hat{r'}$ to $dst$ is $\hat{t'}$, and the path length is $\hat{l'}$,
then $t' > \hat{t'}$, or $t' = \hat{t'}$ and $l' > \hat{l'}$. For either case, because $\hat{p'}$ starts at slice $t$ where $src$ and $r$ are connected, there must be a path $\hat{p} = src + \hat{p'}$ from $src$ to $dst$, which arrives at $dst$ at slice $\hat{t'}$, and the path length is $\hat{l} = \hat{l'} + 1$. Comparing $\hat{p}$ to $p$, we have $t' > \hat{t'}$, or $t' = \hat{t'}$ and $l > \hat{l}$. So, $\hat{p}$ is better than $p$, which contradicts Property 1 that the chosen path $p$ is optimal.

Now that $p'$ is optimal, since \textsc{Routing} selects a single path out of the feasible paths, \textsc{Routing}($r$, $dst$, $t$) may return a different optimal path $\hat{p'}$ equivalent to $p'$, that is $t' = \hat{t'}$ and $l' = \hat{l'}$. Then for the full paths $p$ and $\hat{p}$ from $src$, $t' = \hat{t'}$ and $l = \hat{l}$. So, $\hat{p}$ is also optimal.

Repeating the above proof hop by hop until $dst$, we have hop-wise lookups produce the optimal path.

\par \vspace{-\baselineskip}
\qedhere
\end{proof}

If a packet misses its planned time slice, the switch system adjusts at runtime to reroute the packet by the next available time slice (\S\ref{sec:rerouting}). We prove our runtime adjustment is robust to find the next optimal path starting from the current ToR. 

\vspace{0.3em}\noindent\textbf{\textit{Property 3: Rerouting after missing a planned send time slice gives the next optimal path.}}

\begin{proof}

Let $p$ be the optimal path from $src$ to $dst$, $\{r_i\}$ be the set of intermediate ToRs along $p$, and $\{t_i\}$ be the time slices set for ToR connections of adjacent hops. Assume $t_{i}$ is missed at $r_{i}$ and the current time slice is $t_{c}$ ($t_{c} > t_{i}$). By per-hop lookup, we get a path $p'$ from $r_{i}$ to $dst$ at slice $t_{c}$. According to Property 2, $p'$ is optimal w.r.t. the current time slice $t_c$.

\par \vspace{-\baselineskip}
\qedhere
\end{proof}

\section{\hoho System}\label{sec:implementation}%

In this section, we introduce the implementation of the \hoho algorithm ($\S$\ref{sec:design}) on programmable switches. This implementation later evolved into the \openoptics framework~\cite{OpenOptics, lei2024open}, which provides a general paradigm for deploying various optical \dcn architectures. The on-switch system for \name encompasses key components such as routing lookup ($\S$\ref{sec:lookup_table}), time-based queue management (\S\ref{sec:calendar_queue}), and dynamic adjustment of the empty-queue assumption with packet rerouting ($\S$\ref{sec:rerouting}). Additionally, we address practical considerations ($\S$\ref{sec:practical_issues}) for deploying \name in real-world environments.

\begin{figure}[tbp]
    \centering
    \includegraphics[width=\columnwidth]
    {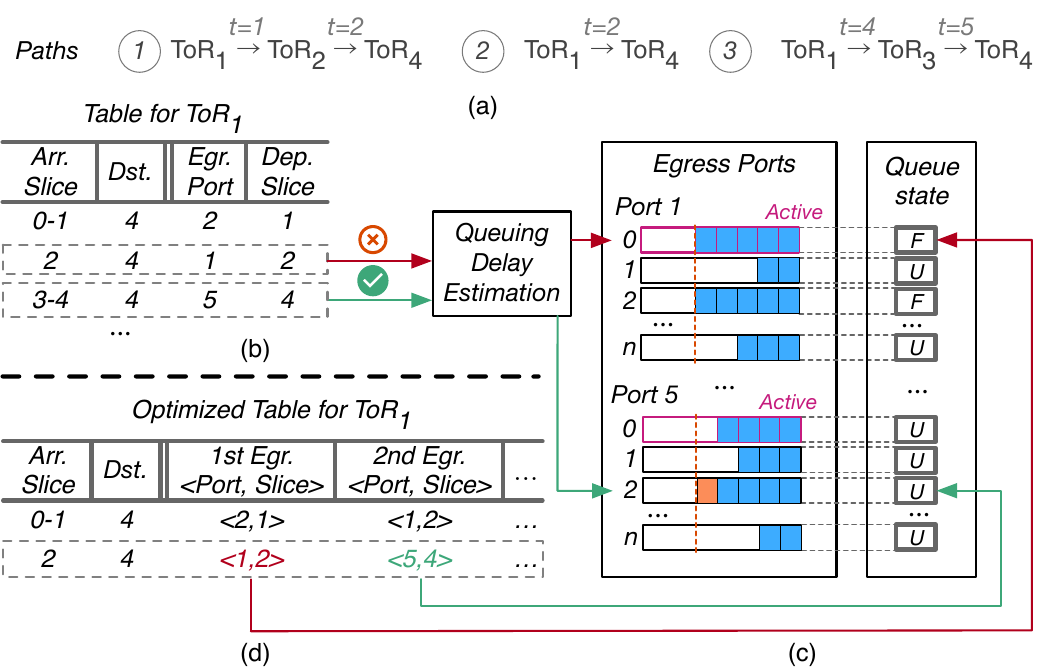}
    \figcap{Illustration of the \hoho switch system ($\S$\ref{sec:implementation}). The current time slice is 2 and the corresponding active queue is 0. (a) \hoho paths ($\S$\ref{sec:alg_design}). (b) $ToR_1$'s lookup table ($\S$\ref{sec:lookup_table}) corresponding to the paths, and queuing delay estimation for packet rerouting ($\S$\ref{sec:rerouting}). (c) Calendar queues for packet buffering ($\S$\ref{sec:calendar_queue}). (d) $ToR_1$'s optimized lookup table for rerouting without packet recirculation ($\S$\ref{sec:rerouting}).
    }\label{fig:tor-example}
\end{figure}

\subsection{Lookup Table}\label{sec:lookup_table}

After the \hoho algorithm computes the full paths for each source-destination ToR pair and for each time slice of packet arrival, i.e., with \textsc{Routing}(\textit{src, dst, t}) ($\S$\ref{sec:alg_design}), we decompose these paths into next-hop lookup tables based on \textit{Property 2} ($\S$\ref{sec:properties}) for ease of implementation in the switch dataplane. The \name lookup table is a simple match-action table where the match fields are the packet's \textit{arrival time slice} and the \textit{destination ToR}, while the lookup (action) data returned consists of the \textit{egress port} and the \textit{departure time slice} when the packet should be transmitted to the next hop (assuming zero queuing).

Existing optical \dcn architectures have provided built-in mechanisms for synchronizing ToRs and hosts with the optical controller~\cite{Sirius, RotorNet, Opera, Corundum}, and \openoptics achieves nanosecond-precision ToR synchronization~\cite{OpenOptics, NOS}.
Therefore, in \name, we pre-load the optical schedule onto ToRs and leverage the synchronization scheme in \openoptics to
determine the arrival time slice of an incoming packet.

As illustrated in Fig.~\ref{fig:tor-example}, the next-hop lookup table for $ToR_1$ in Fig.~\ref{fig:tor-example}b corresponds to the \hoho paths from $ToR_1$ to $ToR_4$ for different time slices in Fig.~\ref{fig:tor-example}a. The first table entry denotes that a packet arriving at $ToR_1$ in time slice \code{t=0} or \code{t=1} will follow path $\textcircled{1}$ and exit in time slice \code{t=1} to $ToR_2$, through egress port \code{p=2} (not shown in the path). A packet arriving in \code{t=0} needs to be buffered until \code{t=1}, while one arriving in \code{t=1} can be sent out immediately. 
Similarly, an incoming packet in \code{t=2} matches the second entry and will be forwarded to $ToR_4$ in \code{t=2} via the direct path \textcircled{2}. 

\subsection{Queue Management}\label{sec:calendar_queue}

In the \hoho lookup table, if the departure time slice is later than the arrival time slice of, the packet must be buffered temporarily for time-scheduled transmission.
The latest programmable switches, e.g., Intel Tofino2, support pausing/resuming of target queues~\cite{lee2020tofino2}. We leverage this feature to enqueue packets meant to be sent out in a later time slice into a designated queue, which we pause until the start of the time slice.
We then resume the queue and keep it active for exactly one time slice before pausing it again.

We realize this design with the calendar queues framework~\cite{CalendarQueue}.
Calendar queues are priority queues, where each queue is associated with a ``calendar day''. 
Packets can be enqueued for a future calendar day depending on their ``rank.''
A \textit{calendar day} is a time slice in our case. 
We form the set of calendar queues using the physical queues per egress port, where we assign each time slice a physical queue sequentially and we wrap around when we have exhausted the available queues.
The \textit{rank} of an ingress packet in our case denotes how many time slices in the future (from the arrival slice) does the packet need to be scheduled for transmission. 
Therefore, we compute the rank of a packet as the \textit{difference} between its \textit{departure time slice} and the \textit{arrival time slice}.

Queue pausing/resuming can be triggered by any packet in the data plane. 
We use an on-chip packet generator~\cite{joshi2019timertasks} to reliably send queue control packets into the data plane at a fixed interval.
We set this interval equal to the \textit{time slice duration}.
Each ToR keeps track of the \textit{active queue} for the current time slice, which is the same across all egress ports. 
\textit{Queue rotation} is triggered by the control packets every time slice interval to pause the current active queue and resume the one for the next time slice.

Fig.~\ref{fig:tor-example}c exemplifies two sets of \textit{calendar queues} for egress ports \code{p=1} and \code{p=5}. Suppose the current time slice is \code{t=2}, and the \textit{active queue} for \code{t=2} is queue \code{q=0} for all the ports. 
An incoming packet in the current time slice (\code{t=2}) will match the second entry (Fig.~\ref{fig:tor-example}b) and get mapped to \code{q=0} of \code{p=1}, because the departure time slice is the same as the arrival/current time slice, and the packet should be enqueued to the active queue to be sent out immediately. 
One time slice later, i.e., \code{t=3}, queue rotation moves the active queue for each port to \code{q=1}. An incoming packet at this time will match the third entry and be mapped to \code{q=2} of \code{p=5}. This is because the departure time slice \code{t=4} is one time slice later than the current time slice \code{t=3}, resulting in a rank of \code{4-3=1}, placing the packet one queue away from the current active queue \code{q=1}.

\subsection{Rerouting}\label{sec:rerouting}

Recall that the \hoho algorithm assumes empty queues (\S\ref{sec:design}). However, at runtime, the ToR system must consider actual queuing delays to determine whether a packet can be delivered within its scheduled time slice. According to \textit{Property~3} ($\S$\ref{sec:properties}), in this case, the packet can be deferred to the next time slice to reroute to the next optimal path.

We have developed a queuing delay estimation scheme that predicts queuing delay of incoming packets in the ingress pipeline before they are enqueued. This method, detailed in the \openoptics paper~\cite{OpenOptics}, achieves an estimation accuracy within \uns{50}, i.e.,  less than one MTU-sized packet at \uGbps{200}.

For an incoming packet, we first lookup the departure slice (Fig.~\ref{fig:tor-example}b) which determines the departure calendar queue (\S\ref{sec:calendar_queue}) for which we estimate the queuing delay.
If the departure queue is a future queue for which the estimated queuing delay is greater than the slice duration, then we consider the queue to be ``full'' and we need to reroute the packet to the next optimal path. 
This is because the future queue when resumed is going to be active only for the slice duration during which the current packet won't be transmitted.
Similarly, in another scenario, if the departure queue is the current active queue, then it is ``full'' if the estimated queuing delay is greater than the remaining time of the current slice.
This is the case with the packet at \code{t=2} in Fig.~\ref{fig:tor-example}b that intends to be enqueued to \code{q=0} of \code{p=1} which is full.

Na\"ively, this rerouting can be achieved by recirculating the packet and re-matching it to a later time slice with a different path, as if the packet arrived at a later time slice.
Since recirculation costs extra pipeline processing and incurs a microsecond-scale delay, we design a \textit{one-shot lookup} mechanism to avoid recirculation.
The main challenge here is that due to the hardware restrictions of the programmable switch pipeline,
the states for queuing delay estimation
can only be accessed once by each packet.
This limitation necessitates packet recirculation to check queue availability for the next optimal path(s). 
We address this challenge by introducing a \textit{queue state table} shown in Fig.~\ref{fig:tor-example}c. 
We maintain the Full (F) or Unfull (U) state for each calendar queue and encode these states into a bit array stored in a single register, which allows us to retrieve all queue states at once.

We also optimize the \hoho lookup table (shown in Fig.~\ref{fig:tor-example}d) by combining the optimal path and the next best paths into a single lookup entry by concatenating their \textit{<egress port, departure time slice>} tuples in the action field. 
In our example, the second entry in the optimized table (Fig.~\ref{fig:tor-example}d) contains the optimal path as in the original table as well as the second best path available at a later time slice. 
We can incorporate more sub-optimal paths to provide higher guarantees that a packet will avoid congestion and find a path during run time.
As Table.~\ref{tab:resource_verify} shows, Tofino2 switch can support combining 10 alternative paths without significant SRAM consumption.

As shown in this example, for each packet, we first retrieve the set of possible paths in a single match and also the Full/Unfull queue state bit array.
We then check the departure queue's state in the order of the paths, e.g., \code{q=0} of \code{p=1} for path <\code{p=1}, \code{t=2}> followed by \code{q=2} of \code{p=5} for path <\code{p=5}, \code{t=4}> in Fig.~\ref{fig:tor-example}d.
In this example, \code{q=2} of \code{p=5} is Unfull and therefore the packet (shown in orange) takes the second best path.
The packet is dropped if no feasible queue can be found. We observe no packet drop in our evaluation ($\S$\ref{sec:evaluation}).

\subsection{Practical Issues}\label{sec:practical_issues}

\para{Packet reordering.} Packet reordering is a common occurrence in optical \dcns, due to frequent topology changes and the corresponding path updates. This phenomenon is particularly acute for \vlb routing, where packets are randomly sprayed across multiple paths of unstable latency, and it is also present in single-path routing approaches such as Opera.
In theory, with the zero queuing assumption, \hoho would avoid packet reordering since packets are sent along the fastest path.
In practice, as in Opera, rerouting and queueing delays can still cause reordering near circuit reconfiguration time.
Generally, modern transport in \dcns can sustain packet reordering in order to support packet spraying~\cite{NDP, Homa}.  
For more extreme cases (i.e., corruption loss), packet reordering detection methods have been proposed specifically for optical \dcns~\cite{TDTCP}.

\para{Transport protocol.}
Design of transport protocols for optical \dcns is an active research area.
For example, reTCP~\cite{reTCP} and TDTCP~\cite{TDTCP} were proposed recently for slowly reconfigured optical \dcns, and Opera is coupled with the low-latency NDP protocol~\cite{NDP}. The sub-\owd time slices \hoho needs to cover requires a more responsive feedback loop. We find Bolt~\cite{Bolt} a good fit for \hoho with the early-feedback control and thus make it the default transport protocol for \hoho.
We found that for \hoho, Bolt outperforms TDTCP, NDP, and DCTCP. 
Interestingly, we also observed that NDP handles congestion poorly under high traffic loads in Opera, while Bolt performs much better. Therefore, in our evaluation ($\S$\ref{sec:evaluation}), we also implement Bolt for Opera for fair comparison.

\para{Co-existence with elephant flows.}
\hoho is primarily designed to minimize latency for mice flows. As discussed in $\S$\ref{sec:prior_routing}, \vlb achieves near-optimal throughput for elephant flows, and its implementation is general to different time slice durations, though the latency expands with longer slices. Hence, in \hoho, we offload elephant flows to \vlb, especially for short time slices. This strategy mirrors Opera's approach, but instead of using a fixed cutoff of $\uMB{15}$ to differentiate mice and elephant flows, we believe the cutoff should be adjusted based on the time slice duration. Intuitively, shorter time slices impose more rerouting overhead on \hoho but reduce the \fct degradation for \vlb, making it more desirable to offload more traffic to \vlb.

We define the \fct slowdown metric $\alpha$ to determine the cutoff flow size for a specific time slice duration. Given the slice duration $u$, we assume an empty network without congestion and derive the maximum \fct of flows\footnote{For a flow of size $x$, we position the flow across all source-destination ToR pairs and calculate the maximum \fct with the \hoho/\vlb algorithm.} as a function of the flow size $x$, denoted as $f(x)$ and $g(x)$ for \hoho and \vlb, respectively. The slowdown function $h(x)=\frac{g(x)}{f(x)}$ presents the acceptable level of \fct slowdown for offloading flows from \hoho to \vlb. Setting $h(x)=\alpha$ produces the cutoff flow size for slice duration $u$. 
We found that $\alpha$ is insensitive in the range between 1.4 to 1.7 under production traffic. In our system, we set $\alpha$ to 1.5. For example, with $\alpha = 1.5$, the cutoff flow size is $\uMB{5}$ for $\umus{2}$ slices and $\uMB{13}$ for $\umus{5}$ slices.

\para{Failure handling.}
The \hoho algorithm is inherently resilient to failures, as failures are analogous to missing scheduled time slices, and rerouting can effectively circumvent such disruptions.
As shown in Fig.~\ref{fig:failure_recovery}, with a 10\% link failure rate, the connectivity loss is limited to 1.56\% ToR pairs when considering a single best path.  
This loss decreases to 0.22\% when considering three best paths. 
While rerouting does defer packets to later time slices, potentially increasing routing latency, Fig.~\ref{fig:fail_fct} demonstrates that this results in only a 12\% degradation in \fcts under typical \dcn failure levels.

Multi-path \hoho can further enhance fault tolerance and we leave this exploration as future work.

\section{Prototype Testbed}\label{sec:testbed}

We implemented the \hoho system ($\S$\ref{sec:implementation}) on Tofino2 switches, which we extended later into the more comprehensive \openoptics framework~\cite{OpenOptics} to support diverse optical \dcn architectures. 
In our implementation, the packet processing capacity of Tofino2 switches allows for a minimum time slice duration of $\umus{2}$ --- the shortest duration achieved by commodity switches known to date. This duration is general enough for most proposed optical architectures as listed in Table~\ref{tab:architecutures}. The derivation of the limit is detailed in the \openoptics paper~\cite{OpenOptics}.

In this section, we validate the correctness of our implementation on a small-scale testbed,
by showing end-to-end \hoho performance with applications and evaluating the switch resource consumption across various optical \dcn scales up to 1024 ToRs.

\para{Testbed setup.}
We setup our testbed to mimic \textit{a flat topology with eight \tors, one host per \tor}, which is the minimum amount to generate a valid Opera schedule with.
We virtualize two Intel Tofino2 programmable switches into four logical \tors each, then a third Intel Tofino switch emulates the circuit switched fabric that interconnects the logical \tors.
Four servers, each equipped with a Mellanox ConnectX-6 Dx dual-port NIC, make eight virtual hosts by splitting the NIC interfaces into separate namespaces.
Inter-\tor uplinks are capped at $\uGbps{10}$, while \tor-host downlinks run at $\uGbps{100}$, which allows us to emulate an oversubscription scenario.

\para{Application performance.}
We now analyze the \hoho{}'s end-to-end application performance. 
We use \textit{Memcached}~\cite{Memcached} to generate mice flows.
Essentially, we run 7 Memslap~\cite{Memslap} benchmarking clients to request $\uKB{4}$ of data (via a \texttt{PULL} operation) continually from a Memcached server; the clients and the server each run on a different host.
We use \textit{iPerf}~\cite{iPerf} to generate elephant flows (i.e., perennial flows with infinitely backlogged data) between hosts in neighboring ToRs.
We vary the time slice duration between $\umus{2}$ and $\umus{50}$, and also realize \vlb and Opera for performance comparison.

\begin{figure}[t]
    \centering
    \begin{minipage}[t]{0.49\columnwidth} \centering
    \includegraphics[width=\linewidth]{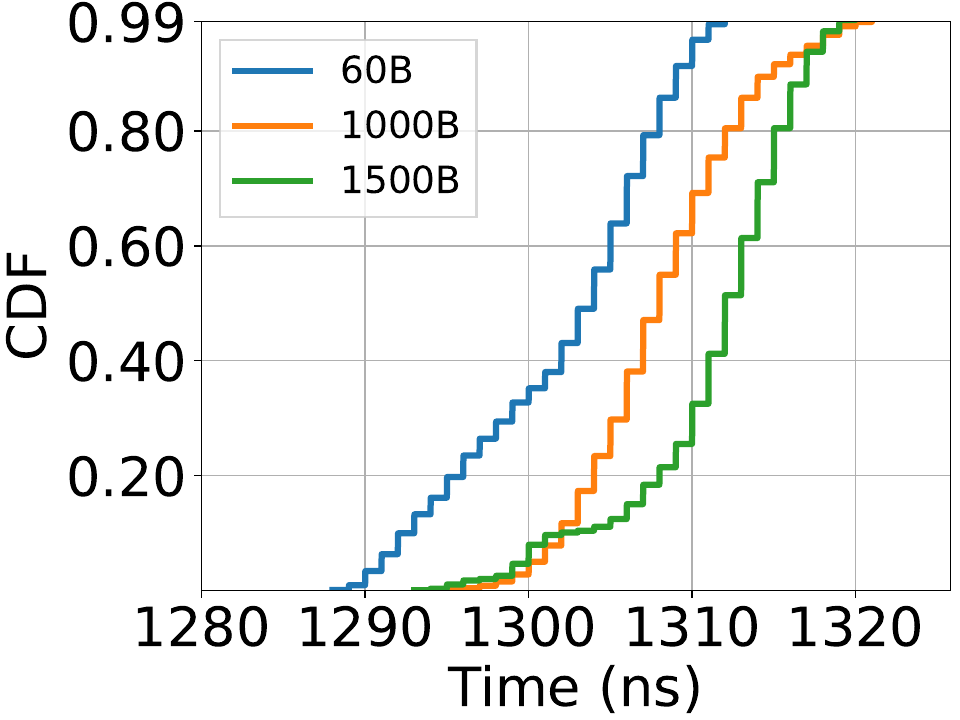}
    \figcap{ToR-to-ToR delay with different packet sizes.}\label{fig:testbed_guardband}
    \end{minipage}
    \hfil %
    \begin{minipage}[t]{0.49\columnwidth}     \includegraphics[width=\linewidth]{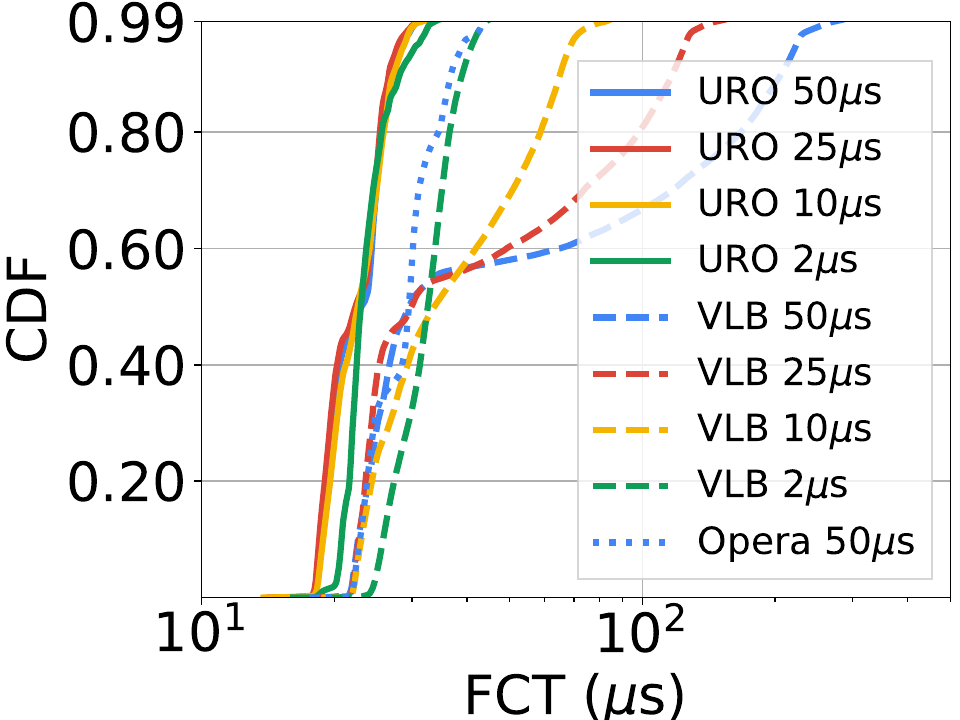}
    \figcap{Memcached \fcts under various time slices.}\label{fig:memcached}
    \end{minipage}
\end{figure}

The \fct distributions of the Memcached flows (Fig.~\ref{fig:memcached}) show that \hoho and \vlb are compatible with different time slice durations, but Opera only functions under $\umus{50}$ slices, which are longer than the OWD.
\hoho achieves significantly lower \fcts than \vlb and Opera:
In the median, \name offers 27.91\% (23.20\%) lower FCTs than VLB (Opera).
In the tail, the \vlb FCTs are as long as the optical cycle, and typically much longer than those of \hoho.
\hoho delivers consistently low \fcts across various time slice durations; we observe a slight increase for the shortest $\umus{2}$ slices due to a reduced duty cycle under the fixed guardband duration of $\uns{200}$.

\begin{table}[tbp]
\footnotesize
\centering
\tabcap{Switch resource usage across optical \dcn scales.%
}
\begin{tabular}{cccccc}
\toprule
($N$, $d$) &Max \#Q/port &   \#Entries/\tor &SRAM  \\ %
\midrule
(108, 6)            & 18          &  1.9K  & 1.13\%       \\ %
(324, 12)           & 27          &  8.8K   & 2.31\%     \\ %
(768, 24)             & 32       & 24.6K      & 6.13\%        \\ %
(1024, 32)              & 32      & 32.8K      & 7.31\%         \\ %
\bottomrule
\end{tabular}%
\label{tab:resource_verify}
\end{table}

\para{Switch resouce usage.}
We take the logical ToRs as a subset of nodes in larger-scale optical \dcns and populate the \hoho table onto them.
Table~\ref{tab:resource_verify} details the switch resource consumption, where $N$ and $d$ represent the number of ToRs and the number of uplinks per ToR, respectively. In the second column, the maximum number of calendar queues per egress port equals $N/d$, essentially the number of time slices per optical cycle.
It remains relatively stable as $N$ and $d$ scale simultaneously for sustained capacity. A 1024-ToR setting 
requires only 32 queues per port, significantly below the capacity of commodity switches~\cite{BFC,SQR}. Tofino2 switches, for instance, support up to 128 queues per port. As shown in Fig.~\ref{fig:max_calendar_q} of our simulation ($\S$\ref{sec:evaluation}), the actual number of queues in use is smaller.
In the last two columns, for a 1024-ToR optical \dcn, \hoho requires 32.8K entries per ToR, which are stored in the switch's SRAM. The low SRAM usage indicates that our per-hop lookup ($\S$\ref{sec:lookup_table}) is efficient and sustainable.

\section{Evaluation}\label{sec:evaluation}%

\begin{figure}[t]
    \begin{minipage}[t]{1\linewidth}
    \centering
        \begin{minipage}[t]{0.49\linewidth}
        \includegraphics[width=\linewidth]
        {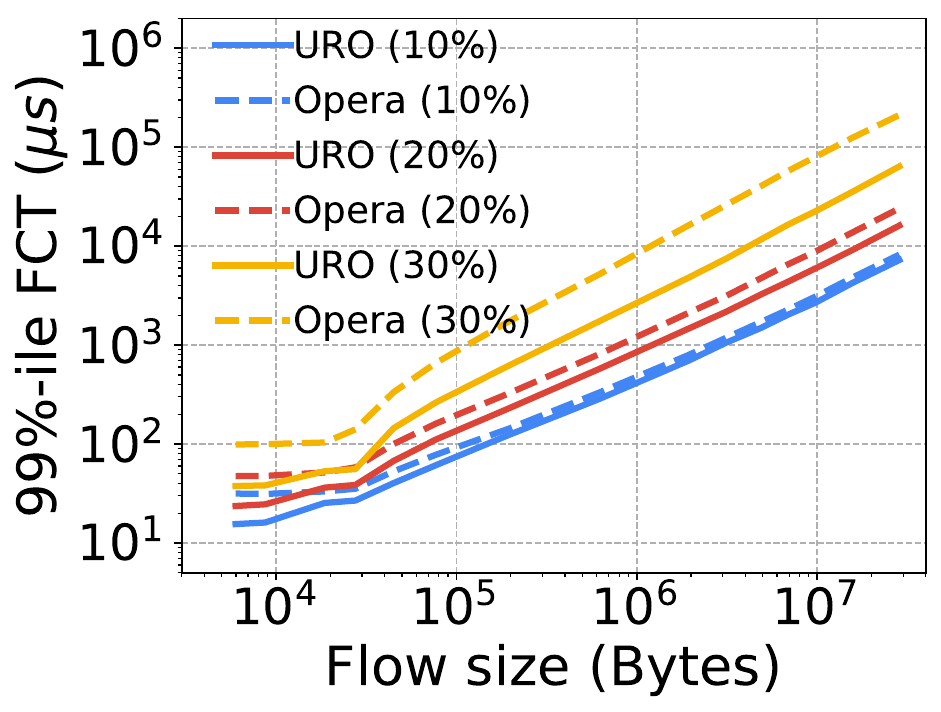}%
        \sfigcap{}\label{fig:hoho_opera_fct_web}
        \end{minipage}%
        \hfill
        \begin{minipage}[t]{0.49\linewidth}
        \includegraphics[width=\linewidth]
        {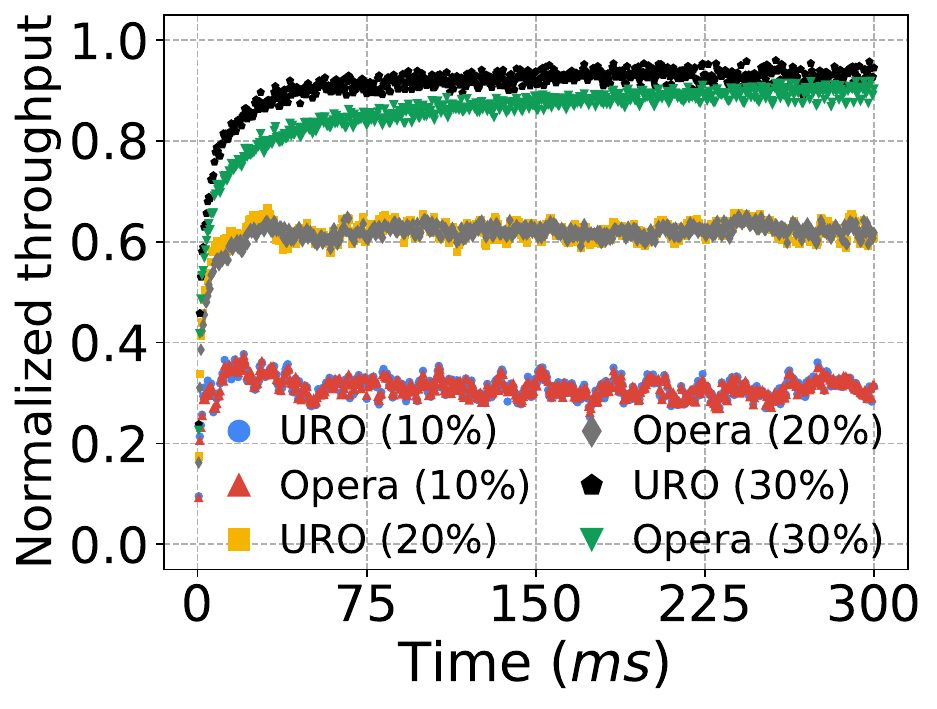}%
        \sfigcap{}\label{fig:hoho_opera_util_web}
        \end{minipage}%
    \figcap{(a) \fcts and (b) throughput comparison with Opera under web search trace.}\label{fig:hoho_opera_fct_util}
    \end{minipage}
\end{figure}

\begin{figure}[t]
    \begin{minipage}[t]{1\linewidth}
    \centering
        \begin{minipage}[t]{0.49\linewidth}
        \includegraphics[width=\linewidth]
        {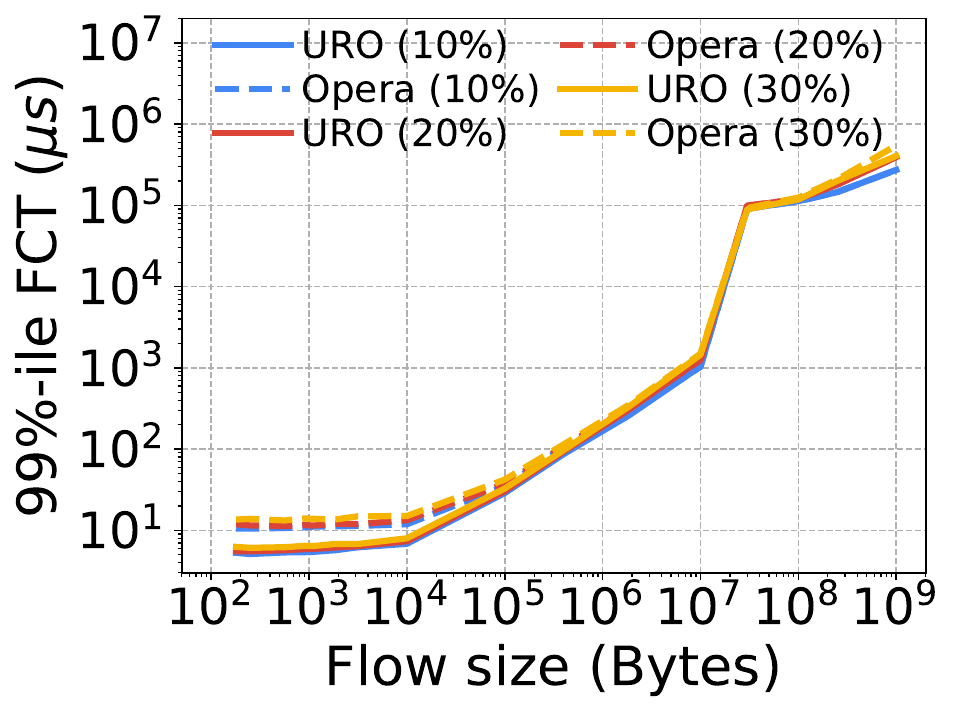}\vspace{-0.0in}\subcaption{}\label{fig:hoho_opera_fct_data}
        \vspace{-0.0in}
        \end{minipage}%
        \hfill
        \begin{minipage}[t]{0.49\linewidth}
        \includegraphics[width=\linewidth]
        {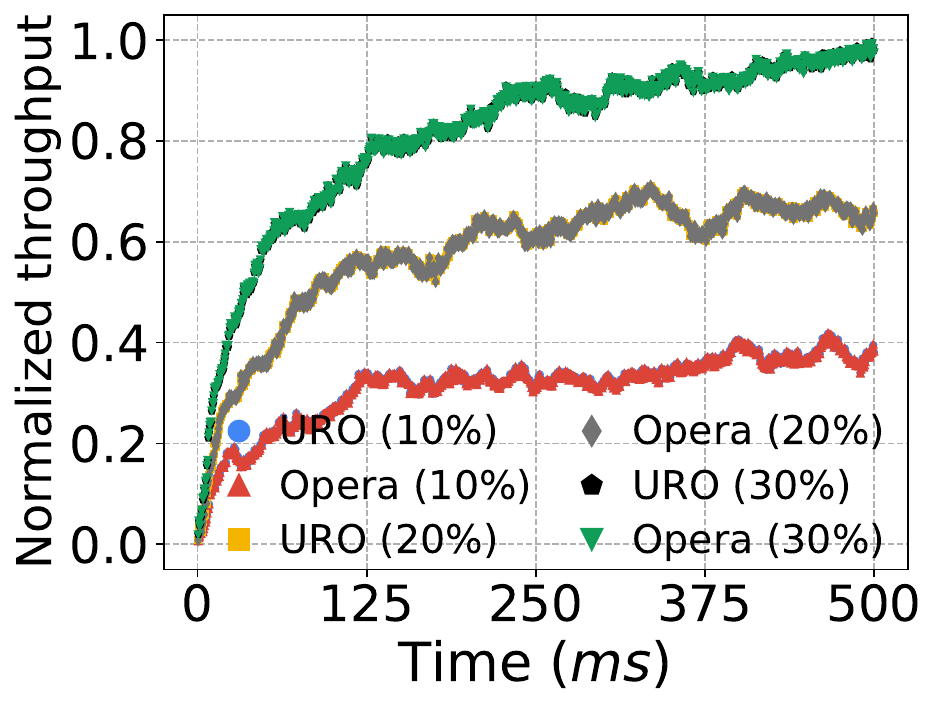}\vspace{-0.0in}\subcaption{}\label{fig:hoho_opera_util_data}
        \end{minipage}%
    \caption{(a) \fcts and (b) throughput comparison with Opera under data mining trace.}\label{fig:hoho_opera_fct_util_data}
    \end{minipage}
\end{figure}

\begin{figure}[t]
    \begin{minipage}[t]{1\linewidth}
    \centering
        \begin{minipage}[t]{0.49\linewidth}
        \includegraphics[width=\linewidth]{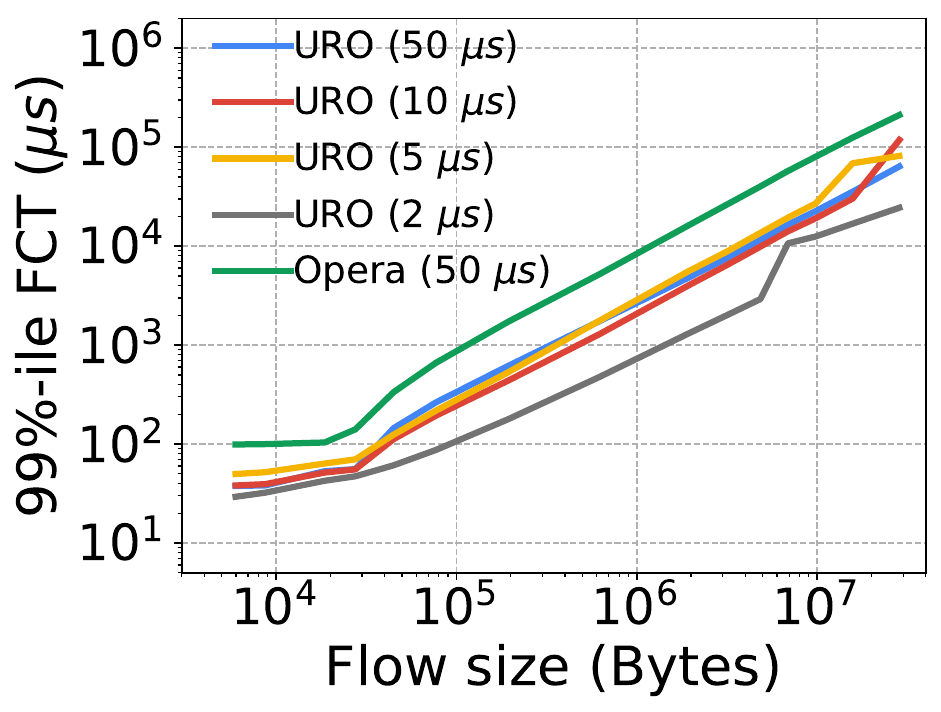}%
        \sfigcap{}\label{fig:hoho_opera_changing_slice_web}
        \end{minipage}
        \hfill
        \begin{minipage}[t]{0.49\linewidth}
        \includegraphics[width=\linewidth]{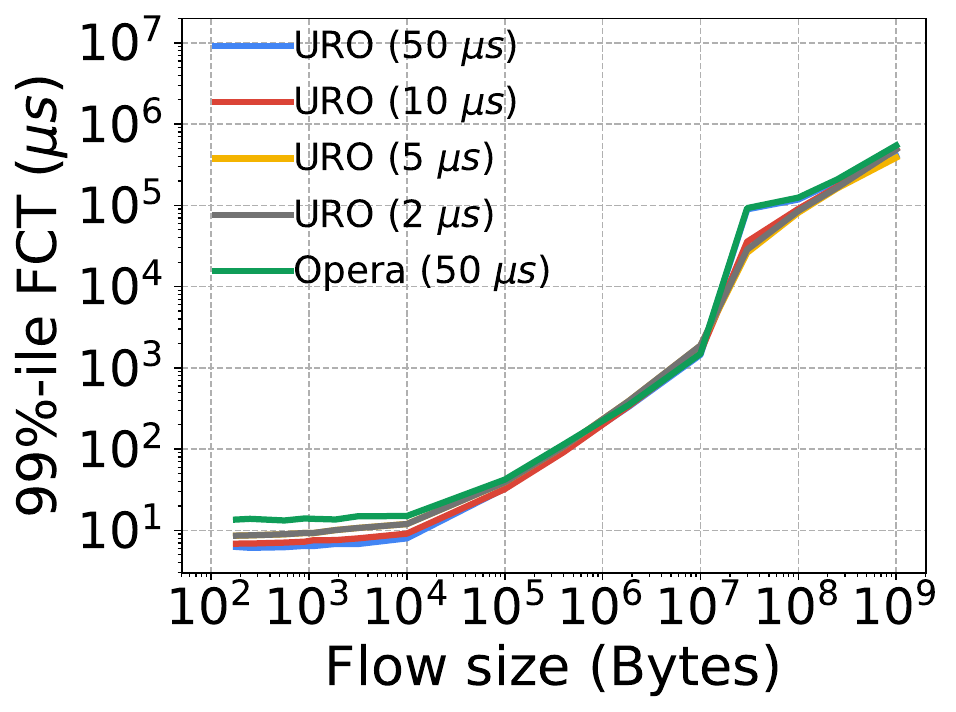}%
        \sfigcap{}\label{fig:hoho_opera_fct_data_30perc}
        \end{minipage}%
    \figcap{\fcts comparison with Opera under different time slices under (a) web search trace and (b) data mining trace.}\label{fig:hoho_opera_fct_fct}
    \end{minipage}
\end{figure}

\begin{figure}[t]
    \begin{minipage}[t]{1\linewidth}
    \centering
        \begin{minipage}[t]{0.49\linewidth}
        \includegraphics[width=\linewidth]{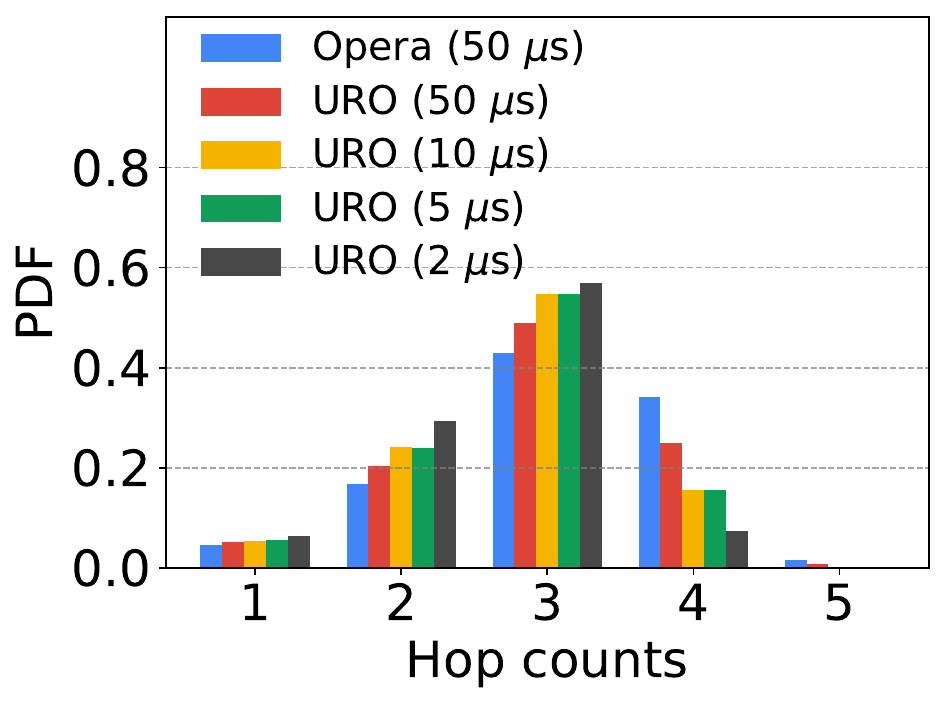}
        \sfigcap{}\label{fig:hop_diverse_slice}
        \end{minipage}
        \hfill
        \begin{minipage}[t]{0.49\linewidth}
        \includegraphics[width=\linewidth]{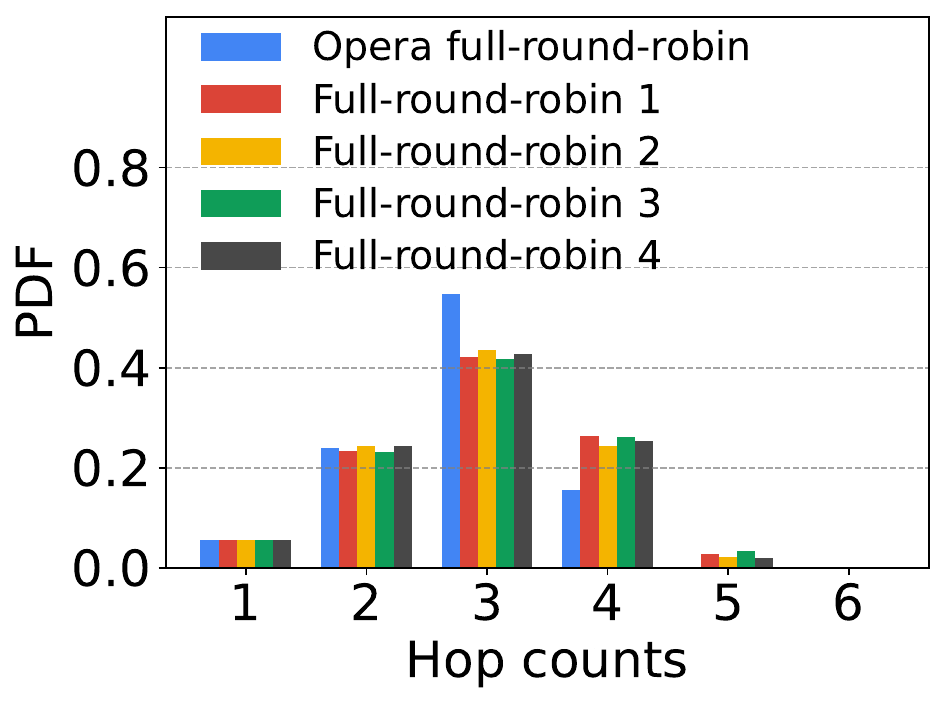}
        \sfigcap{}\label{fig:hop_diverse_schedule}
        \end{minipage}%
    \figcap{(a) Path lengths compared to Opera. (b) Path lengths under diverse schedules.}\label{fig:path_statistics}
    \end{minipage}
\end{figure}

\begin{figure*}[t]
    \begin{minipage}[htbp]{1\linewidth}
    \centering
        \begin{minipage}[htbp]{0.197\linewidth}
        \includegraphics[width=\linewidth]{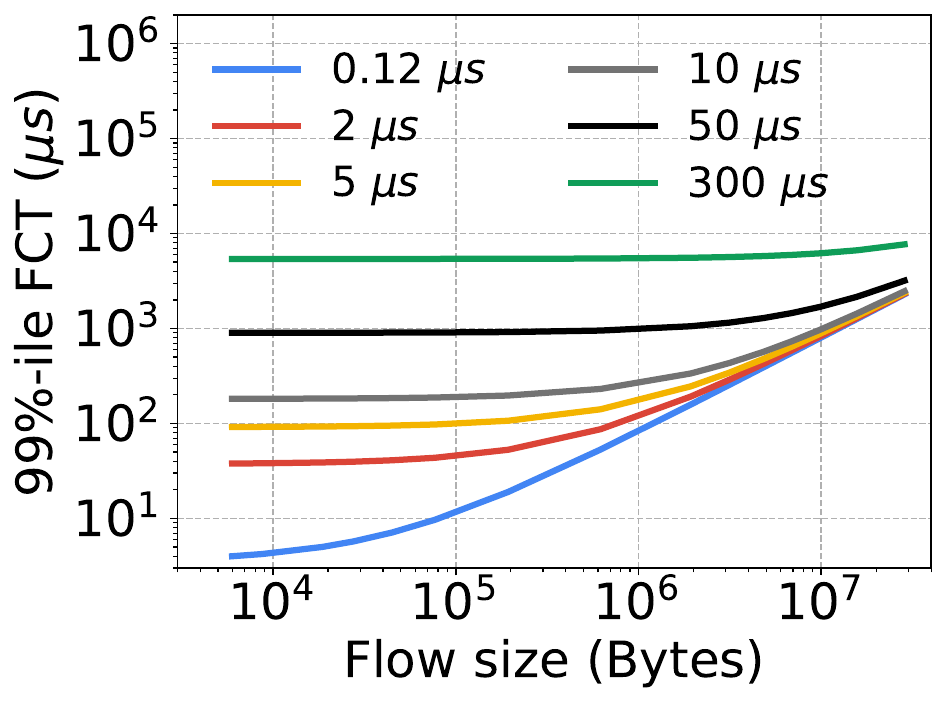}%
        \sfigcap{}\label{fig:sirius_min_fct}
        \end{minipage}
        \hfill
        \begin{minipage}[htbp]{0.197\linewidth}
        \includegraphics[width=\linewidth]{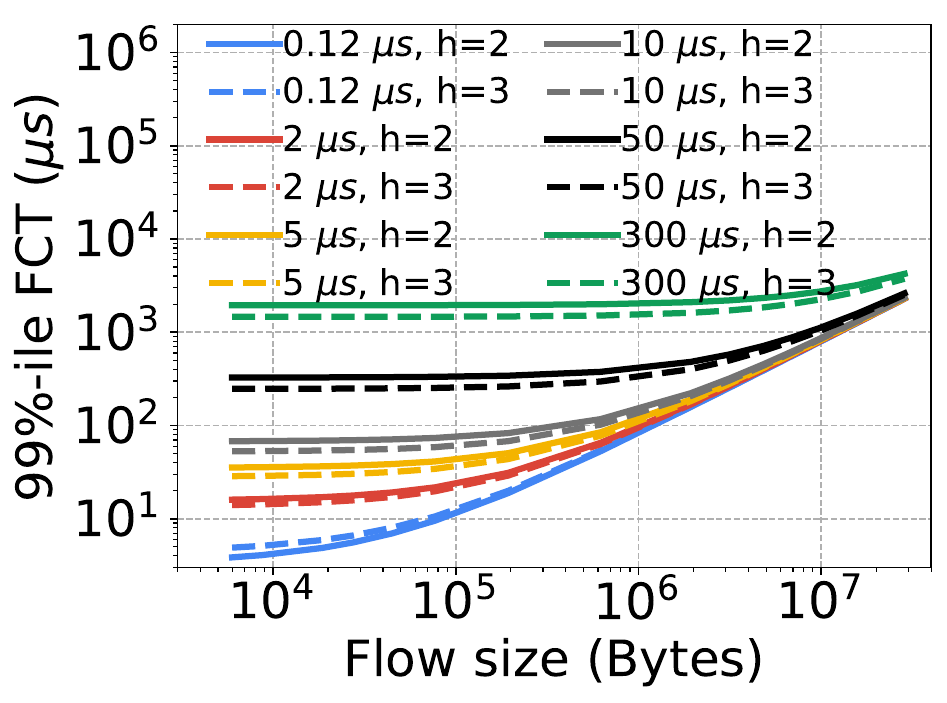}%
        \sfigcap{}\label{fig:shale_min_fct}
        \end{minipage}%
        \hfill
        \begin{minipage}[htbp]{0.197\linewidth}
        \includegraphics[width=\linewidth]{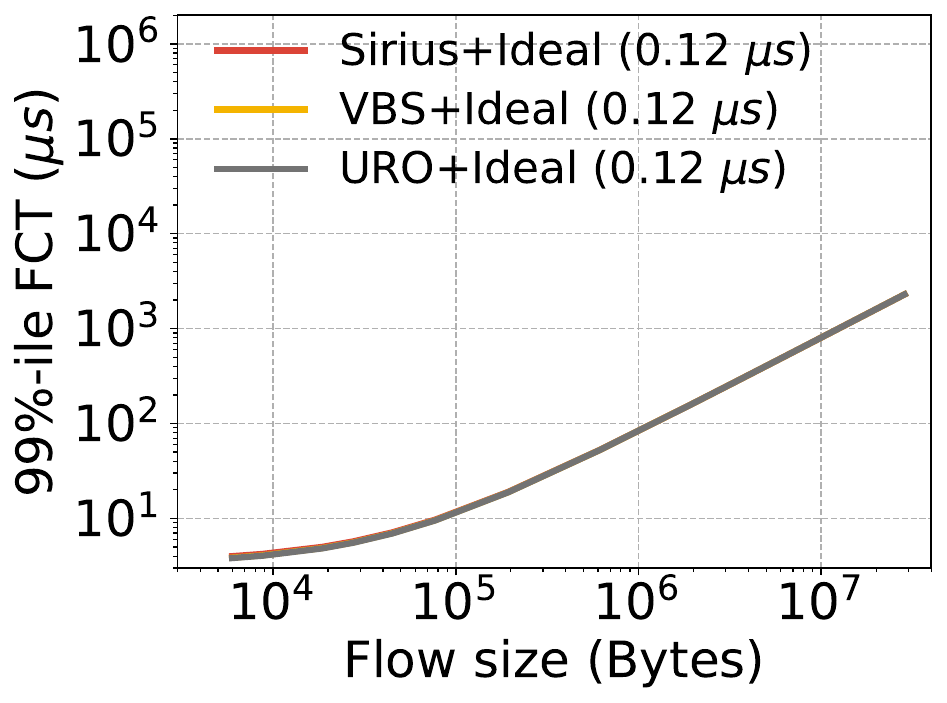}%
        \sfigcap{}\label{fig:all_one_slice_min_fct_0.12us}
        \end{minipage}%
        \hfill
        \begin{minipage}[htbp]{0.197\linewidth}
        \includegraphics[width=\linewidth]
        {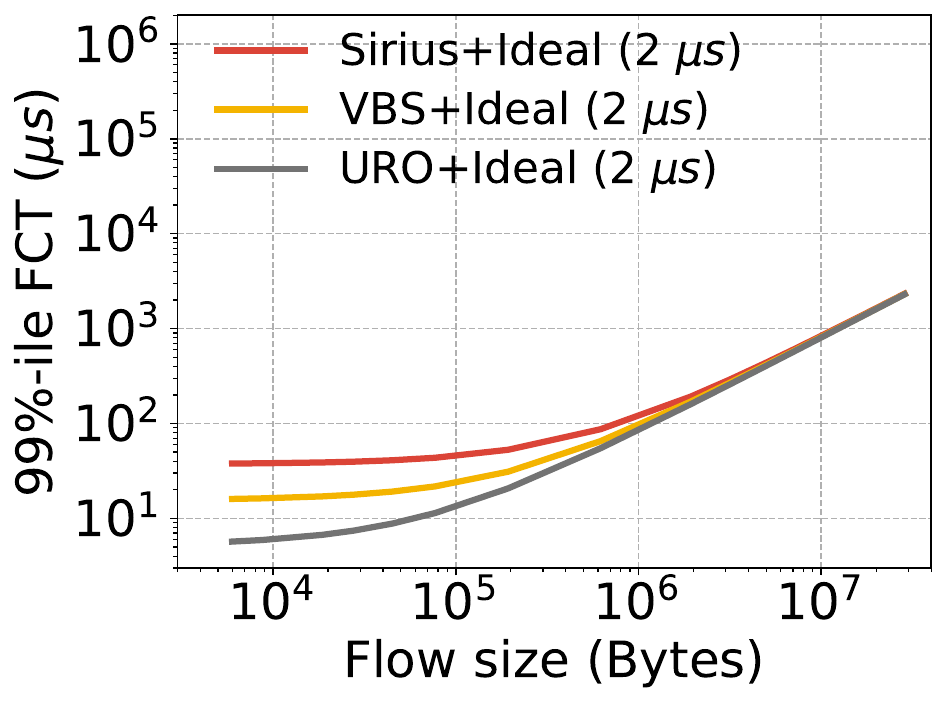}%
        \sfigcap{}\label{fig:all_one_slice_min_fct_2us}
        \end{minipage}%
        \hfill
        \begin{minipage}[htbp]{0.197\linewidth}
        \includegraphics[width=\linewidth]
        {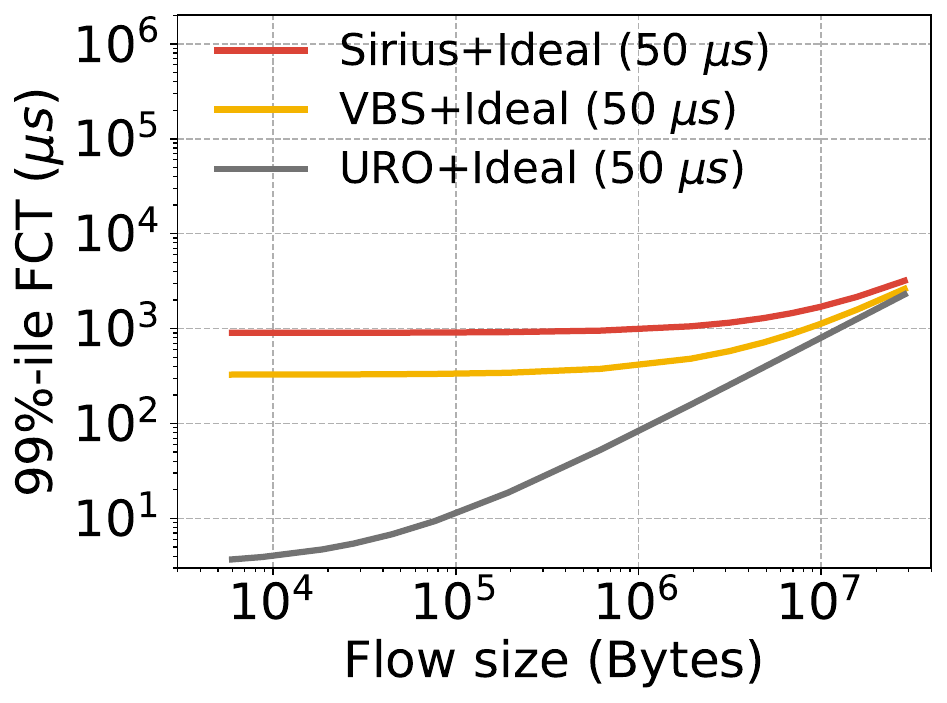}%
        \sfigcap{}\label{fig:all_one_slice_min_fct_50us}
        \end{minipage}%
    \figcap{Theoretical lower-bound \fcts achieved by (a) Sirius and (b) \orn. (c-e) Theoretical lower-bound \fct comparison with Sirius and \orn at $\umus{0.12}$, $\umus{2}$, and $\umus{50}$.}\label{fig:sirius_shale}
    \end{minipage}
\end{figure*}

We now turn to characterizing the performance of \hoho in a large-scale setting using simulations with \dcn traffic. Below, we introduce our experimental setup ($\S$\ref{sec:setup}) and then follow up with the simulations for assessing the different aspects of the \hoho design ($\S$\ref{sec:compare_opera}, $\S$\ref{sec:compare_sirius}, and $\S$\ref{sec:different_settings}).

\subsection{Experimental Setup}\label{sec:setup}

\para{Simulated network.} We implement \hoho on top of the \textit{htsim} simulator, which has been used in prior work to evaluate several routing and transport designs for traditional and optical \dcns~\cite{MPTCP, NDP, Jellyfish, Xpander, Opera}.
We mimic the Opera setup~\cite{Opera} to simulate 108 \tors, where each \tor has 6 downlinks leading to hosts and 6 uplinks leading to 6 \ocses. This results in a 648-hosts network. We set the propagation delay between each \tor pair to $\uns{500}$---approx. 100 meters of fiber---and we consider the propagation delay within a rack negligible.
Differently from Opera, we set link bandwidth to $\uGbps{100}$ to reflect recent changes in \dcn trends.

\para{Baselines.}
It is hard to have direct, apple-to-apple comparisons between \hoho and prior work.
While \hoho is a general, architecture-independent routing solution, our baselines are routing-architecture co-designs that tightly couple a certain type of routing with their optical schedule.
Nevertheless, to put \hoho's performance into context, we compare it with Opera~\cite{Opera}, Sirius~\cite{Sirius}, and \orn~\cite{ORN, ORN_2} on their optical schedule and native routing scheme (refer~$\S$\ref{sec:prior_routing}).
For Opera, we compare in simulations against its optical schedule on both its default time slice duration (at least a \owd) and smaller time slices.
For Sirius (\vlb), we can only simulate microsecond-scale time slices, as that is the minimum duration both our simulator and testbed can support considering realistic network delays on non-customized hardware.
\orn still lacks a system implementation proposal, so it cannot be yet fully evaluated.
We derive theoretical bounds to draw the remaining comparisons with Sirius and \orn.

\para{Transport protocols.}
We run Bolt (as per $\S$\ref{sec:practical_issues}) as the transport for Opera and \hoho, and RotorLB as the transport for \vlb.
We chose to run Bolt on Opera to provide a fair comparison, as we find it to be perform better than \ndp~\cite{NDP}. %

\para{Circuit settings.}
In simulation, we run two different schedules.
The first one is the \textit{Opera schedule} that offsets the reconfiguration times across \ocses to achieve rotating continuous paths (i.e., one \ocs reconfigures at a time), which we use to compare with Opera.
The second one is a \textit{full-round-robin schedule} we get by removing the offsets in the Opera schedule (i.e., all \ocses reconfigure at once), which we use to compare with Sirius and \orn.
While for \hoho and Sirius we simulate with time slices as small as $\umus{2}$ (down to $\umus{0.12}$ for theoretical bounds), Opera by design does not support time slices shorter than a \owd, which is $\umus{50}$ on our setup.

\para{Workloads.} We run the same \textit{web search} and \textit{data mining} traces from Microsoft's production \dcns~\cite{VL2, DCTCP} used to evaluate Opera. 
The web search trace predominantly includes mice flows, mostly under Opera's long flows cutoff. In contrast, the data mining trace contains more elephant flows, with sizes up to $\uGB{1}$ and the majority of packets originating from flows over the cutoff.
We scale these traces to achieve up to 30\% utilization on the host-to-\tor links, which saturates the core bandwidth.

\subsection{Comparison with Opera}\label{sec:compare_opera}\vspace{-0.0in}

We start by evaluating \hoho on the Opera schedule. We show a direct \fct comparison against Opera, as well as  how its better paths and flexibility allow for further improvements when moving away from the Opera's constraints.

\para{\hoho offers lower \fcts.}
In Fig.~\ref{fig:hoho_opera_fct_util}, we compare \hoho with Opera under the same settings by varying the traffic load from 10\% to 30\% web search trace.
\textit{\hoho consistently outperforms Opera}.
Specifically, \hoho achieves 99\textsuperscript{th} percentile \fcts that are 53\% lower at 20\% traffic load, and this improvement becomes even more significant at 30\% traffic load, with 2$\times$ reduction. Regardless of setting, \hoho adopts shorter paths than Opera, reducing congestion in the network core during high traffic load and enhancing performance.
These observations concerning \hoho's performance relative to Opera also hold in the simulations with the data mining traces (Fig.~\ref{fig:hoho_opera_fct_util_data}).

\para{\hoho supports shorter time slices.}
\hoho is all but limited to the default Opera settings.
Fig.~\ref{fig:hoho_opera_fct_fct} shows \hoho \fcts under different time slices at 30\% traffic load compared to Opera.
\textit{Both mice and elephant flows take advantage of \hoho's flexibility}.
Mice traverse a lighter-loaded network thanks to more aggressive offloading to \vlb, while elephants, routed using \vlb, reap the benefits of the smaller time slices to reduce tail latency. 
With a $\umus{2}$ time slice, \hoho achieves a \fct reduction between 1.4$\times$ and 12.8$\times$ compared to Opera for web search traces. For mice flows in data mining traces, \hoho reduces \fct by approximately 50\%, and for elephant flows, the reduction is between 12\% and 45\%.

\para{\hoho has shorter paths.} Fig.~\ref{fig:hop_diverse_slice} illustrates the hop count distributions using the Opera schedule under different time slices. \textit{\hoho achieves shorter paths (fewer hops) compared to Opera}. This benefit is more pronounced as the time slice duration decreases, because for shorter slices, the advantage of waiting at a ToR for the next slice outweighs the delay caused by taking additional consecutive hops. Notably, the slice duration of $\umus{10}$ maintains 99.99\% of paths under 4 hops and 84.39\% under 3 hops. This configuration reduces the average path length from 3.11 hops in Opera to 2.80. Reducing the slice duration further to $\umus{2}$ decreases the average path length even more, to~2.65.

Furthermore, Fig.~\ref{fig:hop_diverse_schedule} displays the hop count distributions for the full-round-robin Opera schedule and four more random full-round-robin schedules. These schedules feature shorter paths compared to the Opera schedule because the underlying expander graphs do not need to guarantee full connectivity at all times.
Opera can operate exclusively on the Opera schedule, whereas \hoho is compatible with any optical schedule.

\para{URO has higher throughput.}
The advantage of using shorter paths, which pay less bandwidth cost, is reflected in the higher throughput of \hoho.
Fig.~\ref{fig:hoho_opera_util_web} shows the throughput over time, normalized to Opera's maximum achievable throughput (32\% ToR-to-downlink utilization).
At 30\% traffic load, when the network is saturated, \textit{\hoho realizes 9.3\% higher throughput than Opera under the same settings}.

\subsection{Comparison with Sirius and \orn
}\label{sec:compare_sirius}\vspace{-0.00in}

Next, we evaluate \hoho on the full-round-robin schedule using the state-of-the-art \vlb baselines. We first discuss the scalability of \vlb routing, which lead us to derive theoretical latency bounds of \hoho, Sirius and \orn. Then, we show the advantage of \hoho over \vlb on sub-OWD schedules.

\para{\vlb has low latency at packet-granularity.}
Sirius and \orn are tightly designed to work with \vlb routing under their packet-granularity optical schedules.
As described in $\S$\ref{sec:prior_routing}, compared to common minimum-latency routing, \vlb uniformly balances traffic in the network at a latency cost.
When time slice duration decreases to the nanosecond level this latency cost is minimal, so the default \vlb routing adopted by these architectures offers an excellent balance of latency and throughput.
In Fig.~\ref{fig:sirius_shale}, we derive the theoretical lower-bound FCTs achieved by Sirius and \orn on their schedule\footnote{The theoretical lower-bound \fcts is derived with an ideal transport protocol in an empty network. The network parameters are detailed in $\S$\ref{sec:setup}.}.
Particularly, Fig.~\ref{fig:all_one_slice_min_fct_0.12us} shows how \textit{at packet-granularity \vlb offers comparable latency to \hoho's minimum-latency routing, while also offering uniform load balancing}.

\para{\vlb latency degrades at microsecond-scale.}
Fig.~\ref{fig:sirius_min_fct} and Fig.~\ref{fig:shale_min_fct} show the theoretical lower-bound FCTs for Sirius and \orn over increasing time slice duration.
\textit{As time slice duration increases, \vlb routing incurs very high latency costs}.
As described in $\S$\ref{sec:prior_routing}, the $h$ parameter in \orn controls the latency-throughput trade-off.
While \orn indeed achieves lower latency than Sirius at higher $h$\footnote{For \orn, setting $h=2$ results in lower \fct compared to $h=3$ because, in such a scaled network, the waiting time for circuits becomes less significant than the propagation and transmission delays.}, it degrades as badly at our network scale.
On the other hand, \hoho still achieves near-ideal latency across larger time slices in the full-round-robin schedule, as shown in Fig.~\ref{fig:all_one_slice_min_fct_2us} and Fig.~\ref{fig:all_one_slice_min_fct_50us}.

\begin{figure}[t]
    \begin{minipage}[htbp]{1\linewidth}
    \centering
        \begin{minipage}[htbp]{0.49\linewidth}
        \includegraphics[width=\linewidth]{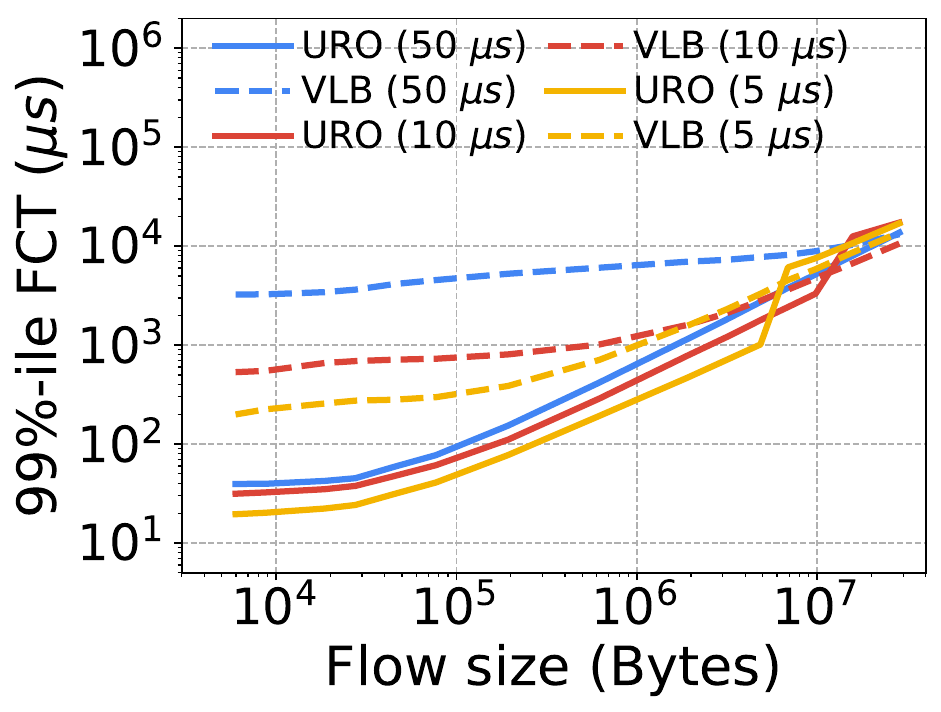}\vspace{-0.0in}\subcaption{}\label{fig:hoho_vlb_fct_web}
        \vspace{-0.0in}
        \end{minipage}%
        \begin{minipage}[htbp]{0.49\linewidth}
        \includegraphics[width=\linewidth]{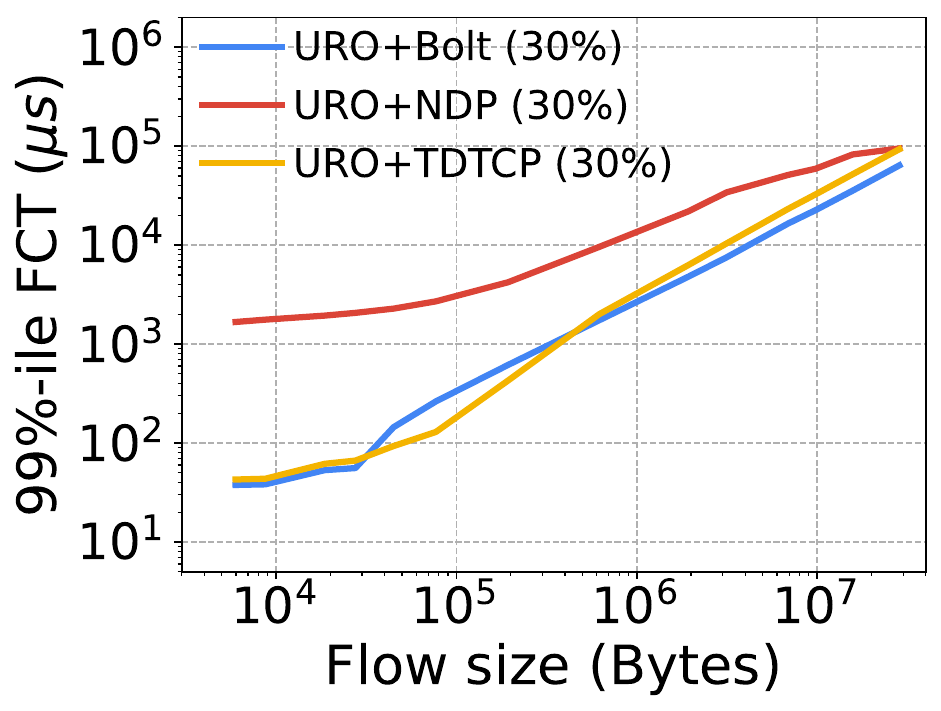}\vspace{-0.0in}\subcaption{}\label{fig:hoho_vlb_fct_data}
        \vspace{-0.0in}
        \end{minipage}
    \caption{\fct comparison with \vlb under (a) web search and (b) data mining trace.}\label{fig:fig:hoho_vlb_fct}
    \end{minipage}
    \vspace{-0.0in}
\end{figure}

\para{\hoho outperforms Sirius at microsecond-scale.}
Finally, we compare \hoho with Sirius in network simulations with time slice durations we can reasonably run on our simulated network ($\S$\ref{sec:setup}).
Fig.~\ref{fig:hoho_vlb_fct_web} shows \fcts for \hoho and Sirius under varying time slice settings with web search trace.
The benefits of \hoho for mice flows are immediately evident, especially as the time slice duration increases.
\textit{Mice flows in \hoho achieve up to two orders of magnitude lower \fct than Sirius at microsecond-scale time slices}, while elephant flows still achieve comparable throughput.
Our direct comparison with Sirius shows how \vlb by itself is insufficient to accommodate both mice and elephant flows in microsecond-level schedules.
In this regard, unlike Opera, \hoho can always provide low-latency routes without the need for altering existing schedules.

Fig.~\ref{fig:hoho_vlb_fct_data} compares the \fcts of \hoho and Sirius under different time slice settings using the data mining trace. This shows a similar performance improvement to what is observed with the web search trace. As the duration of the time slices increases, the advantages of \hoho for mice flows become more apparent, with \hoho achieving up to two orders of magnitude lower FCT than Sirius at microsecond-scale time slices.

\subsection{Scrutiny of Different Aspects
}\label{sec:different_settings}%

\begin{figure}[t]
    \begin{minipage}[htbp]{1\linewidth}
    \centering
        \begin{minipage}[htbp]{0.49\linewidth}
        \includegraphics[width=\linewidth]{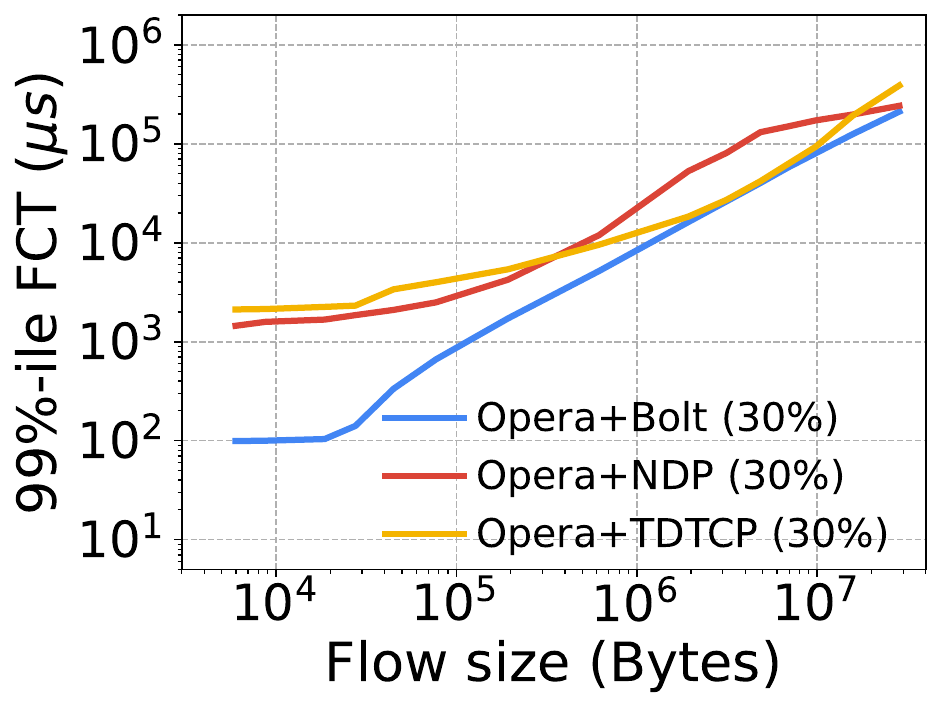}\vspace{-0.0in}\subcaption{}\label{fig:varying_cc_30perc_opera}
        \vspace{-0.0in}
        \end{minipage}%
        \begin{minipage}[htbp]{0.49\linewidth}
        \includegraphics[width=\linewidth]{figures_appendix/varying_cc_30perc_hoho.pdf}\vspace{-0.0in}\subcaption{}\label{fig:varying_cc_30perc_hoho}
        \vspace{-0.0in}
        \end{minipage}
    \caption{\fct performance of Bolt, NDP, and TDTCP with (a) Opera and (b) \hoho.}\label{fig:varying_cc_30perc}
    \end{minipage}
    \vspace{-0.0in}
\end{figure}

\begin{figure}[t]
    \centering
    \begin{minipage}{.49\linewidth}
        \centering
        \includegraphics[width=\linewidth]{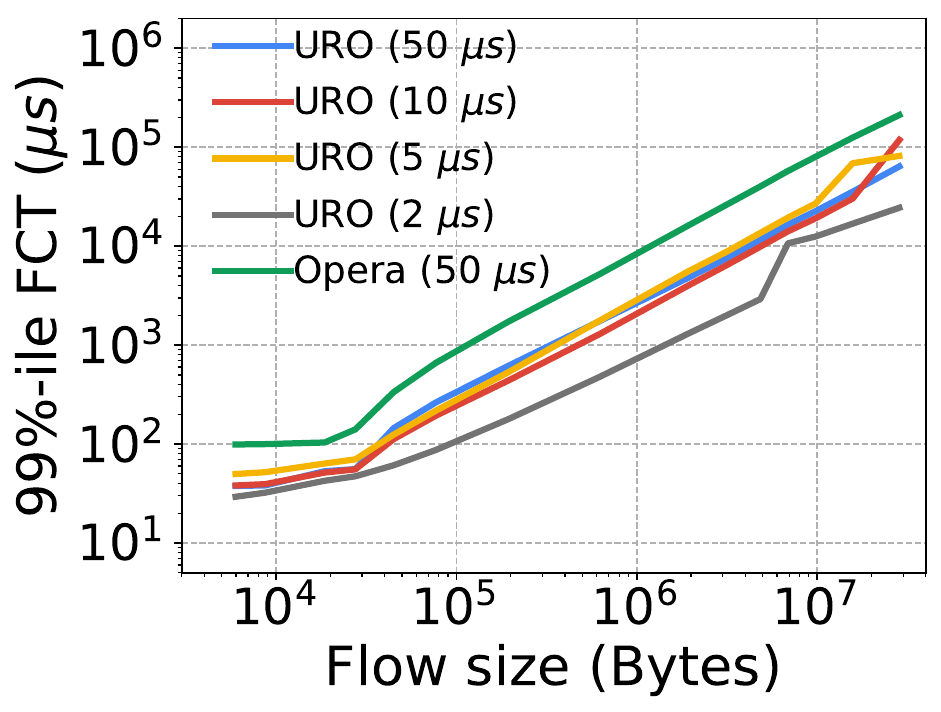}
        \subcaption{}\label{fig:hoho_opera_changing_slice_alpha=1.4}
    \end{minipage}%
    \begin{minipage}{.49\linewidth}
        \centering
        \includegraphics[width=\linewidth]{figures_appendix/hoho_opera_changing_slice_alpha=1.5.pdf}
        \subcaption{}\label{fig:hoho_opera_changing_slice_alpha=1.5}
    \end{minipage}

    \begin{minipage}{.49\linewidth}
        \centering
        \includegraphics[width=\linewidth]{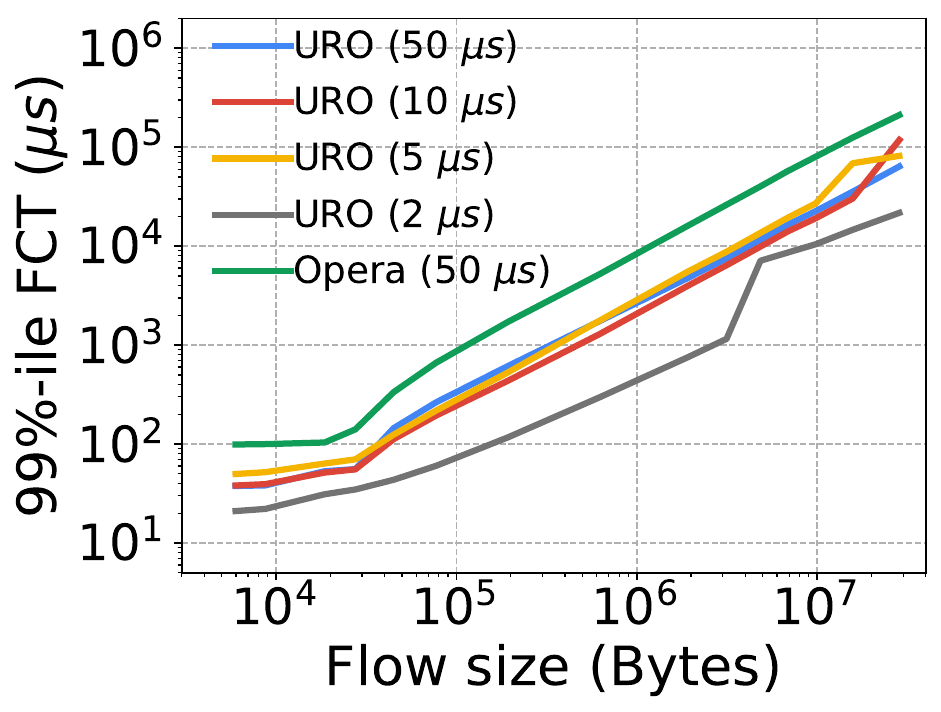}
        \subcaption{}\label{fig:hoho_opera_changing_slice_alpha=1.6}
    \end{minipage}%
    \begin{minipage}{.49\linewidth}
        \centering
        \includegraphics[width=\linewidth]{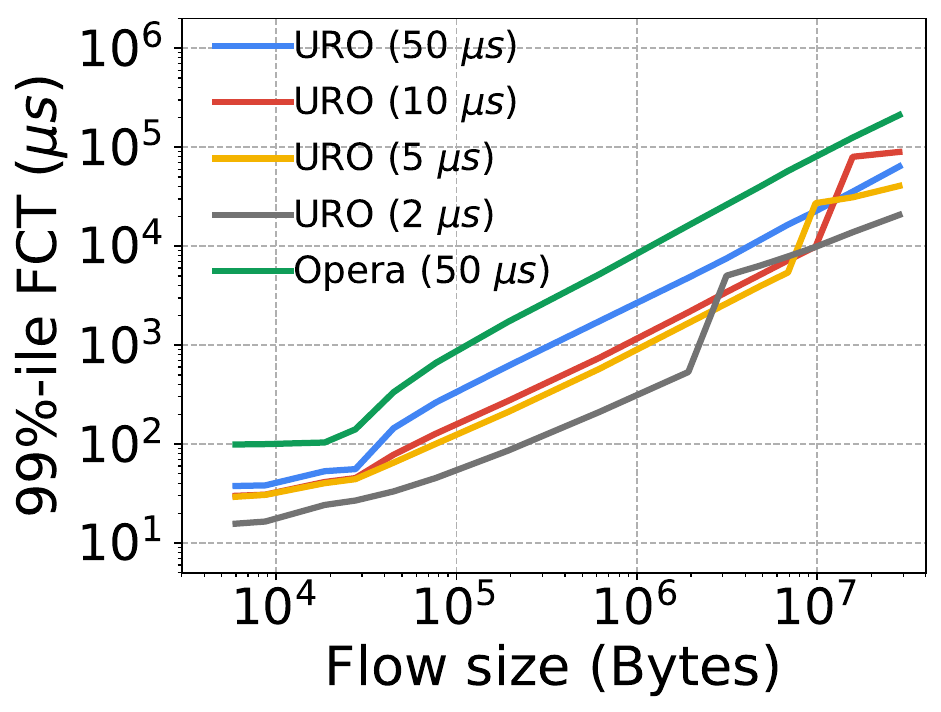}
        \subcaption{}\label{fig:hoho_opera_changing_slice_alpha=1.7}
    \end{minipage}

    \caption{\fcts when setting slowdown metric $\alpha$ to (a) 1.4, (b) 1.5, (c) 1.6, and (d) 1.7.}\label{fig:hoho_opera_changing_slice_alpha}
\end{figure}

\para{Impact by Transport Protocols.}
In order to fairly pick the most fitting transport for both \hoho and Opera, we test three main candidates: NDP \cite{NDP}, TDTCP \cite{TDTCP}, and Bolt \cite{Bolt}.
NDP was originally used for evaluating Opera due to its low-latency properties.
TDTCP has been later proposed as a TCP variation to specifically deal with optical \dcns, but has yet to be evaluated on Opera in particular.
Lastly, Bolt is a very recent contribution that is able to provide sub-RTT feedback, which we believe fitting for the dynamicity of an optical \dcn.
Fig.~\ref{fig:varying_cc_30perc} shows the \fct performance of Bolt, NDP, and TDTCP with 30\% web search traffic load on both \hoho and Opera on the same time slice setting. As we see Bolt consistently outperforming the other schemes, particularly in Opera, we pick it as the default transport for both architectures for a fair comparison.

\begin{figure}[t]
    \centering
    \begin{minipage}[t]{0.49\columnwidth} \centering
    \includegraphics[width=\linewidth]{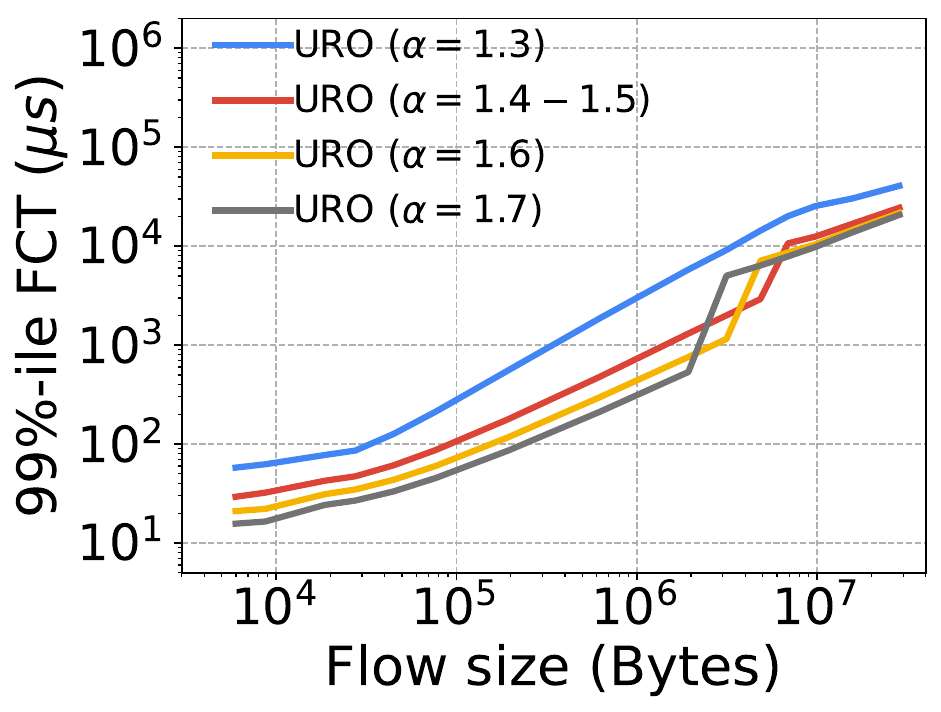}
    \figcap{Impact by slowdown metric $\alpha$.}\label{fig:2us_changing_alpha}
    \end{minipage}
    \hfil %
    \begin{minipage}[t]{0.49\columnwidth}     \includegraphics[width=\linewidth]{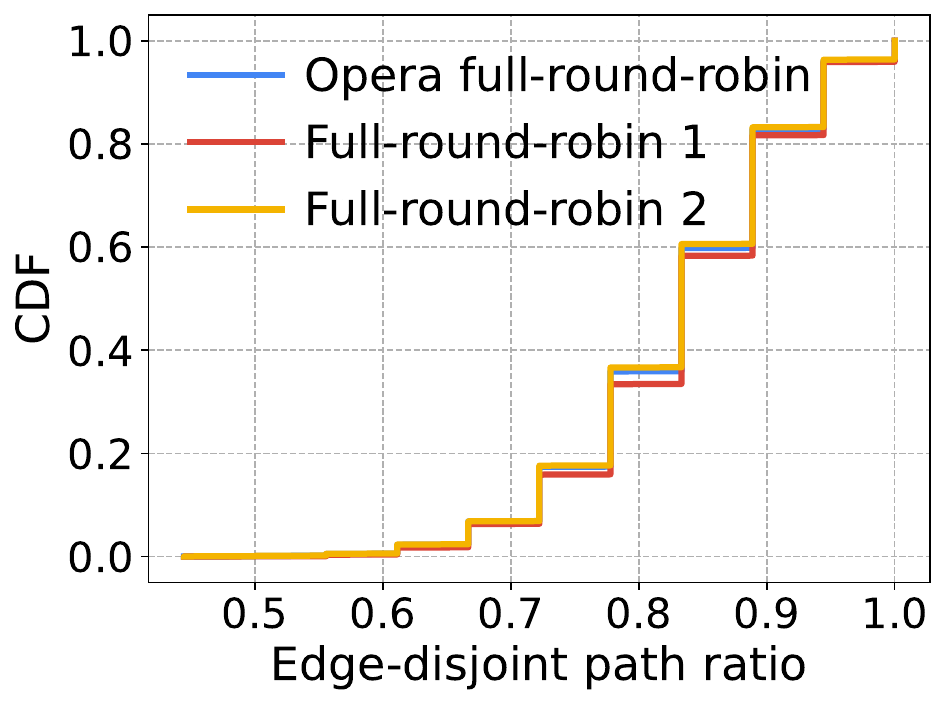}
    \figcap{Edge-disjoint path ratios.}\label{fig:edge_disjoint_cdf}
    \end{minipage}
\end{figure}

\para{Impact of slowdown metric $\alpha$.}
Fig.~\ref{fig:2us_changing_alpha} shows the impact of the slowdown metric $\alpha$ on \hoho FCT for the $\umus{2}$ Opera schedule.
As discussed in $\S$\ref{sec:practical_issues}, higher $\alpha$ will move more flows to the high latency, high throughput \vlb paths.
While it is evident from the figure that setting a lower slowdown will lead to a smaller FCT degradation for flows after the cutoff, we also show that in some cases a more aggressive $\alpha$ can result in an overall performance improvement. 
Fig.~\ref{fig:2us_changing_alpha} in particular shows such a case, where both mice and elephant flows benefit from having $\alpha \ge 1.4$.
For our evaluation, we end up setting $\alpha = 1.5$, as we experimentally find the parameter to be insensitive around that range.

\para{Sensitivity test of slowdown metric $\alpha$.} To assess the sensitivity of the slowdown metric $\alpha$, we vary $\alpha$ from 1.4 to 1.7 and display the \fct results in Fig.~\ref{fig:hoho_opera_changing_slice_alpha}. \hoho consistently outperform Opera across this range, demonstrating its robustness to changes in $\alpha$.

\begin{figure}[t]
    \begin{minipage}[htbp]{1\linewidth}
    \centering
        \begin{minipage}[htbp]{0.49\linewidth}
        \includegraphics[width=\linewidth]{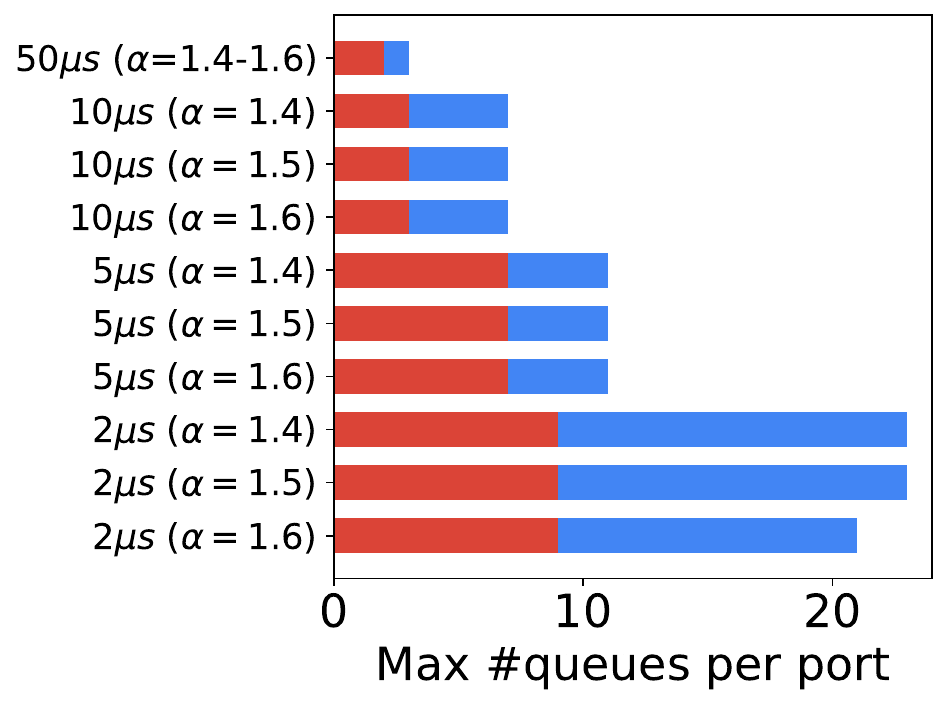}%
        \sfigcap{}\label{fig:max_calendar_q}
        \end{minipage}%
        \begin{minipage}[htbp]{0.49\linewidth}
        \includegraphics[width=\linewidth]{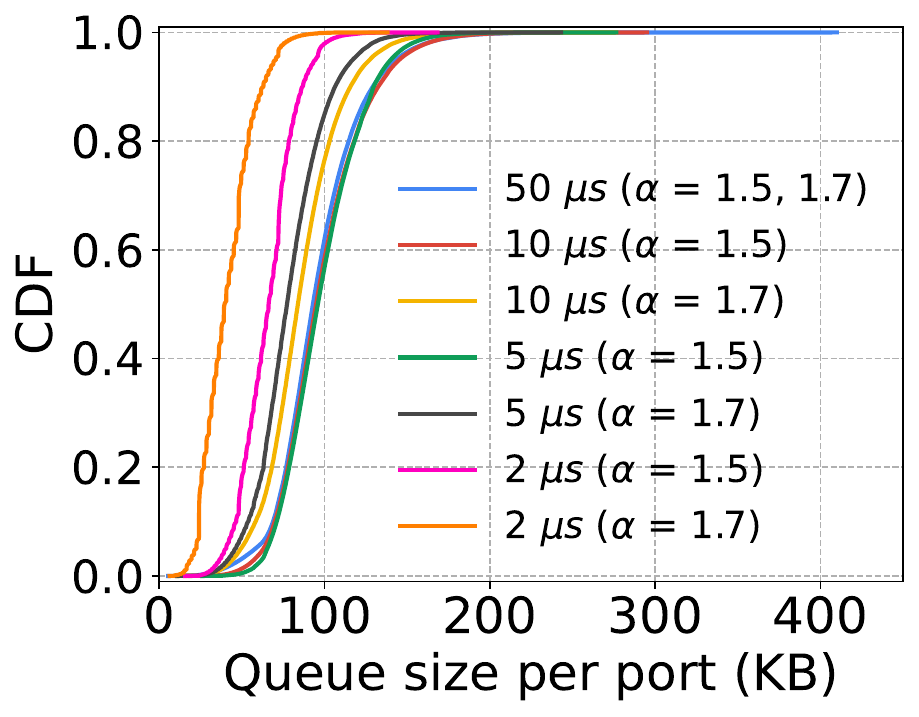}%
        \sfigcap{}\label{fig:q_cdf}
        \end{minipage}
    \figcap{(a) Maximum (blue) and 99\textsuperscript{th} percentile (red) number of queues per port. (b) Queue occupancy per port. The results for $\umus{300}$ are similar to those for $\umus{50}$ and are not included in the figure.}\label{fig:sim_overhead}
    \end{minipage}
\end{figure}

\para{Number of queues.} \hoho in theory could need up to $N$ calendar queues per port---or $N/d$ for the full-round-robin schedule ($\S$\ref{sec:testbed})---to manage mice flows.
In practice, this number is bound by the maximum buffer size at a port even under worst-case traffic.
Fig.~\ref{fig:max_calendar_q} shows both the 99\textsuperscript{th} percentile and the absolute maximum number of concurrently active calendar queues reached in simulation by each setting during stress tests with 30\% traffic load.
We observe that the required number of queues is actually less than 10 in 99.99\% of cases for our most demanding $\umus{2}$, $ \alpha = 1.5$ setting.
This is well within the capabilities of current commodity switch ASICs~\cite{BFC}.

\para{Queue occupancy.} Fig.~\ref{fig:q_cdf} shows the maximum queue occupancy per uplink port sampled every $\umus{500}$ in simulations with 30\% traffic load. \hoho demands the largest buffer sizes with $\umus{50}$ time slice, with the median and tail queue occupancy per port being $\uKB{93}$ and $\uKB{410}$, respectively. A 128-port ToR switch, with half its ports linked to the optical fabric, would require a total buffer size of $\uMB{5.95}$ for the median case.

\begin{figure}[t]
    \begin{minipage}[t]{1\linewidth}
    \centering
        \begin{minipage}[t]{0.49\linewidth}
        \includegraphics[width=\linewidth]{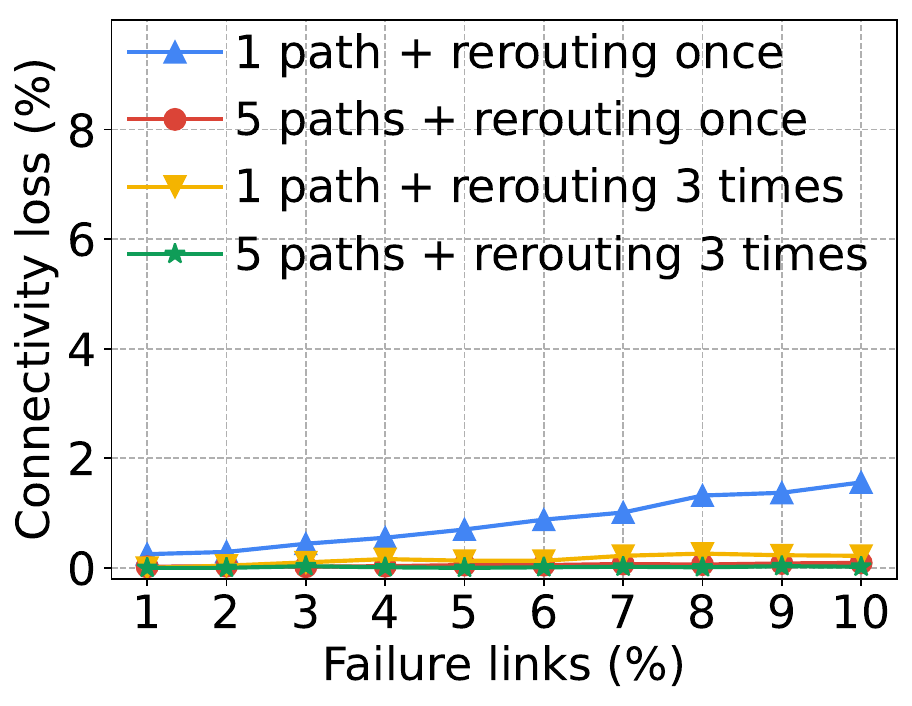}%
        \sfigcap{}\label{fig:fail_port}
        \end{minipage}%
        \hfill
        \begin{minipage}[t]{0.49\linewidth}
        \includegraphics[width=\linewidth]
        {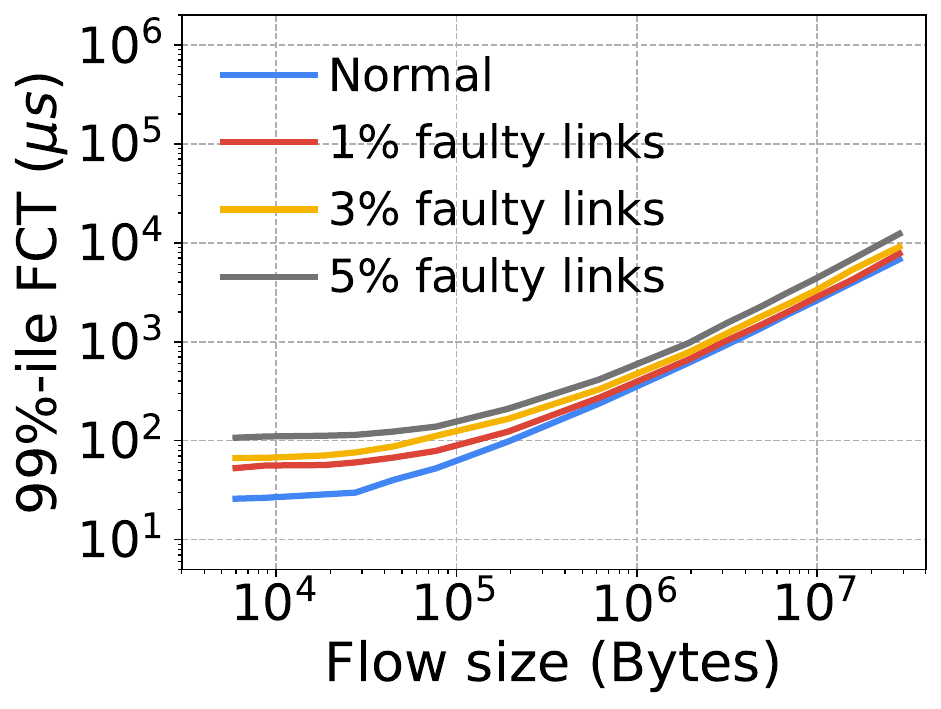}%
        \sfigcap{}\label{fig:fail_fct}
        \end{minipage}%
    \figcap{(a) Connectivity loss under failed links. (b) FCTs under 1\%, 3\%, and 5\% faulty links.}\label{fig:failure_recovery}
    \end{minipage}
\end{figure}

\begin{figure}[t]
    \begin{minipage}[t]{1\linewidth}
    \centering
        \begin{minipage}[t]{0.49\linewidth}
        \includegraphics[width=\linewidth]{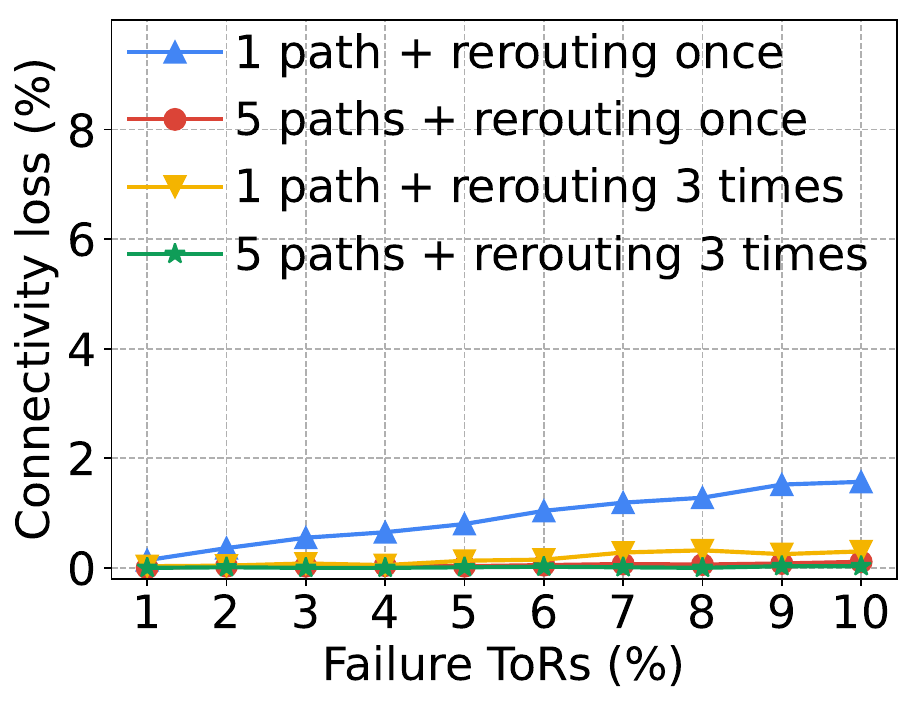}\vspace{-0.0in}\subcaption{}\label{fig:fail_tor}
        \vspace{-0.0in}
        \end{minipage}
        \hfill
        \begin{minipage}[t]{0.49\linewidth}
        \includegraphics[width=\linewidth]
        {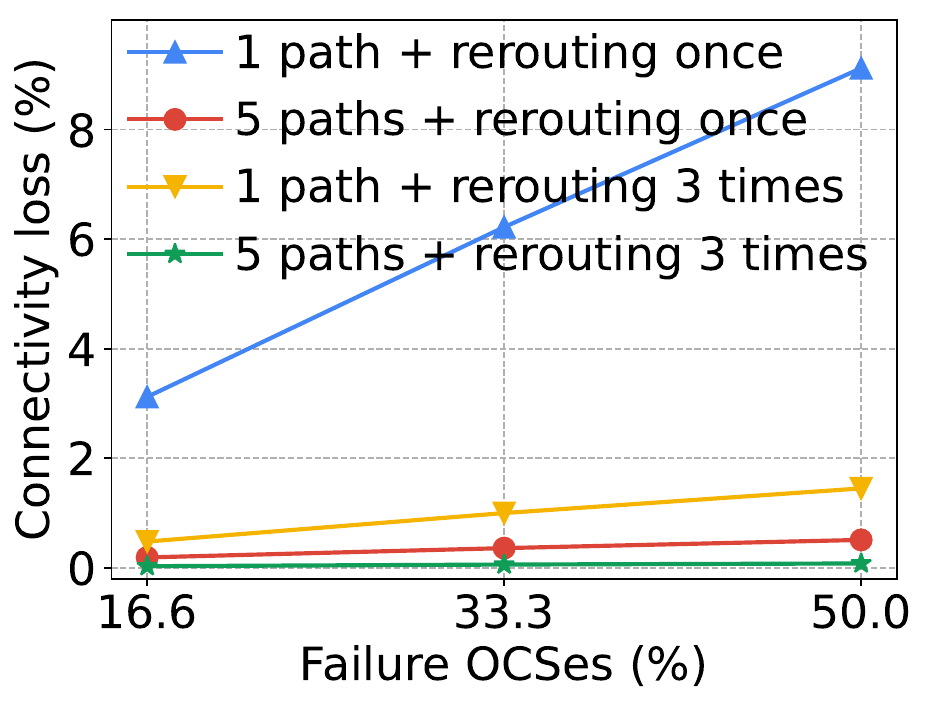}\vspace{-0.0in}\subcaption{}\label{fig:fail_ocs}
        \vspace{-0.0in}
        \end{minipage}%
    \caption{Connectivity loss under failed (a) ToRs and (b) OCSes.}\label{fig:fail_additional}
    \end{minipage}
\end{figure}

\para{Failure recovery.} Fig.~\ref{fig:fail_port} illustrates \hoho's resilience to failures. With a 10\% link failure rate, connectivity loss is restricted to 1.56\% ToR pairs with a single best path. 
This loss decreases to 0.22\% with three
best paths. 
Additionally, using five paths and only one rerouting further reduces the loss to just 0.09\%. 
Recall that we proposed one-shot lookup optimization ($\S$\ref{sec:rerouting}) to avoid recirculation while choosing subsequent best paths, so the overhead is minimal for rerouting multiple times.
In the figure we show that rerouting more times can reduce the connectivity loss.
In Fig.~\ref{fig:fail_fct}, \hoho exhibits low \fct degradation under up to 5\% link failures. We ensure zero connectivity loss in these simulations by not limiting the number of rerouting attempts. Considering we show the 99\textsuperscript{th} percentile \fcts, the \fct degradation for most flows is moderate.

Fig.~\ref{fig:fail_additional} shows the connectivity loss when there are failures in the ToRs and OCSes. When 10\% of the ToRs fail, the connectivity loss is 1.57\% after one rerouting attempt. This loss reduces to 0.30\% after three rerouting attempts. For OCS failures, a 16.6\% failure rate results in a connectivity loss of 3.12\% after one rerouting attempt, which decreases to 0.48\% after three attempts. These results demonstrate the robustness of the system against various types of failures.

We further examine the new path selected during rerouting to explain \hoho's robustness against failures. The new selected path is called edge-disjoint if it uses different ports from the path in the previous time slice. We calculate the ratio of edge-disjoint paths to total paths for each ToR pair and plot the distribution in Fig.~\ref{fig:edge_disjoint_cdf}. The figure shows that over 80\% of the ToR pairs have an edge-disjoint path ratio exceeding 0.7. A higher ratio indicates a higher probability that \hoho can circumvent faulty links by switching to a new path in the next time slice, confirming \hoho's robustness against failures.

\section{Related Work}\label{sec:related_work}%

\para{Optical \dcn architectures.}
There is a large body of work regarding architectural designs for slow-switched~\cite{MegaSwitch, ProjecToR, Firefly, Mordia, JupiterEvolving} and fast-switched~\cite{Opera, RotorNet, Sirius, Shoal}  optical \dcns.
\hoho is not an optical \dcn architecture proposal per se, but a general routing scheme for fast-switched architectures.
As shown in $\S$\ref{sec:evaluation}, \hoho can function on a wide range of time slices.

\para{Optical schedules.}
Fast-switched optical \dcn architectures run alongside a predetermined optical schedule, such as the simple round-robin schedule adopted by Sirius~\cite{Sirius}. Optimized schedules have been proposed to improve network connectivity and ensure routing properties. Particularly, as discussed in $\S$\ref{sec:prior_routing} and analyzed in $\S$\ref{sec:compare_sirius}, Opera~\cite{Opera} forms constrained expander graphs to guarantee continuous paths, and \orn~\cite{ORN, ORN_2} introduced multi-dimensional round-robin schedules to lower the tail latency. Besides, Mars~\cite{Mars} proposed a schedule that optimizes throughput under buffer constraints, a common case for intermediate nodes. \name is general to different optical schedules, including these unorthodox ones.

\para{Optical routing.}
We have discussed extensively about routing approaches for fast-switched optical \dcns in $\S$\ref{sec:prior_routing}.
An early version of \name was introduced in HOHO~\cite{HOHO}. It 
presented the basic concept of identifying the fastest paths but did not compare its approach to state-of-the-art solutions such as Sirius and \orn. 
Additionally, it provided only a high-level system outline without developing a prototype testbed, limiting the ability to validate its proposed methods and assess practical feasibility. In contrast, we conducted comprehensive evaluations on \hoho with different workloads, transport protocols, and varying time slice durations. We implemented a prototype and validated \hoho's feasibility, demonstrating its effectiveness in optimizing path selection and reducing latency under various conditions. \hoho also addressed practical issues and failure handling in the context of deployment. UCMP~\cite{UCMP} designed a multi-path routing solution to balance throughput and latency. It focused on the theoretical design and can be seen as an ECMP equivalent for optical \dcns. \name is orthogonal in proposing a new routing paradigm general to different optical hardware architectures. We emphasized on realizing the paradigm on programmable switches with a simple algorithm minimizing routing latency.

\para{Optical transport.}
reTCP~\cite{reTCP} and TDTCP~\cite{TDTCP} are TCP extensions proposed for fast-switched optical \dcns with super-OWD time slices, which still allow network feedback to travel end-to-end. References~\cite{de2024poster, de2024rethinking} review existing transport protocols in optical \dcns and advocate for an opportunistic credit-based protocol. \name facilitates exploration of transport protocols also for sub-\owd slices, which is more challenging due to discontinuous paths but are essential for performance.

\section{Conclusion}\label{sec:conclusion}%
In this paper we presented \hoho, a unified, practical solution across different time slice durations and optical schedules.
We have demonstrated its performance benefits through testbed and simulation experiments.
Nonetheless, \hoho also raises some questions and opens up problems in the space of optical \dcn designs.
First of all, \hoho optimizes for latency on various time slices, while \vlb for throughput.
On packet-granularity time slices, \orn finds the optimal trade-off between throughput and latency, but this optimization space is largely unexplored for microsecond-scale time slices.
Secondly, while we experimentally chose Bolt as the default transport protocol for \hoho, we still believe it is far from optimal in sub-OWD scenarios.
This work calls for further research on sub-OWD transport protocols, that may also need to account for discontinuous paths. Lastly, \hoho could open up the design space for new sub-OWD optical schedules, which were previously bound to either direct-path or VLB routing.

\bibliographystyle{IEEEtran} 
\bibliography{paper}

\end{document}